# Soft lamellar solid foams from ice-templating of self-assembled lipid hydrogels: organization drives the mechanical properties

Niki Baccile,[a*] Ghazi Ben Messaoud,[a,†] Thomas Zinn,[b] Francisco Fernandes[a,*]

[a] Sorbonne Université, Centre National de la Recherche Scientifique, Laboratoire de Chimie de la Matière Condensée de Paris, LCMCP, F-75005 Paris, France

[†] Current address: DWI- Leibniz Institute for Interactive Materials, Forckenbeckstrasse 50, 52056 Aachen, Germany

[b] ESRF - The European Synchrotron, 71 Avenue des Martyrs, 38043 Grenoble, France

* Corresponding authors: niki.baccile@sorbonne-universite.fr; francisco.fernandes@sorbonne-universite.fr

**Abstract**

Ice-templating, also referred to as freeze-casting, is a process exploiting unidirectional crystallization of ice to structure macroporous materials from colloidal solutions. Commonly applied to inorganic and polymeric materials, we employ it here to cast soft self-assembled matter into spongy solid foams. Use of ice-templating to cast soft matter is generally confined to polymers. In the case of polymeric hydrogels, cross-linking ensures a good stability towards the harsh conditions (fast cooling at temperatures as low as -80°C) employed during ice-templating. However, freeze-casting of soft systems held together by weak interactions, like in physical gels, has not been explored, because the nonequilibrium conditions could easily disrupt the nano and macroscale organization of self-assembled matter, resulting in a cruel loss of mechanical properties. Whether this is a general assumption or a more specific relationship exists between the structure of the physical gel and the properties of the macroporous solid after ice-templating is the question addressed in this work. We compare two self-assembled lipid hydrogels, of analogous chemical composition and comparable elastic properties under ambient conditions, but different structure: isotropic entangled self-assembled fibers against heterogeneous lipid lamellar phase. Our results show that both materials possess the same phase (fibrillar and lamellar) before and after freeze-casting but the mechanical properties are absolutely at the opposite: the fibrillar hydrogel provides a brittle, highly anisotropic, macroporous fibrous solid while the lamellar hydrogel provides soft, spongy, solid foam with isotropic Young moduli of several kPa, in the same order of magnitude as some soft living tissues.

**Introduction**

Processing of matter from solution to materials commonly occurs under nonequilibrium conditions, often implying mechanical stress alongside with rapid variations of physicochemical parameters like temperature, ionic strength, and relative humidity, just to cite the most important ones. Soft complex fluids[1] are governed by weak, reversible, interactions. Their structural properties are thus highly sensitive to small variations in environmental conditions. For this reason, the process of casting them into materials, that relies on highly energetic processes such as evaporation or solvent crystallization, has a crucial impact on the ability to obtain self-supported materials. Under these circumstances, the nanoscale structure of the fluid is critical as it will determine the ability to withstand the constraints applied by the casting process.

Ice-templating has emerged as one of the most versatile processing techniques to design macroporous materials from solutions and suspensions.[2] The process is based on the directional growth of ice crystals through an aqueous solution (or suspension), leading to the segregation of the solutes (or suspended particles) from the newly formed ice crystals. Owing to the limited solubility of most solutes in ice — hexagonal ice is known to form a limited number of solid solutions with few compounds such as some strong acids, nitrates or alkali hydroxides at very low molar ratios[3,4] — the technique enables for a progressive concentration of the solutes in the interstitial space formed between ice crystals. Most applications of ice templating focus on shaping ceramic green bodies in presence of binding water-soluble polymers to obtain lightweight ceramics[5], for the elaboration of macroporous biomaterials[6] or even to fabricate highly efficient thermal insulating materials.[7]

The simplicity in implementing freeze casting, coupled with the ability to promote controlled segregation between solutes and frozen water, is a powerful tool to shape and introduce porosity in soft materials. However, the physicochemical conditions associated to freeze-casting are extreme, as they imply the use of fast temperature variation, strong temperature gradients and reaching temperatures as low as -80°C. Last but not least, ice expansion during freezing is believed to apply strong anisotropic pressure to the newly segregated solute. These harsh conditions are compatible with soft cross-linked polymer systems,[8,9] but they may not be with those soft fluids, of which the stability depends on the subtle equilibrium among intermolecular forces, entropic gain or on physical phenomena like entanglements, in the case of physical gels. For these reasons, the number of examples in which freeze-casting process has been applied to soft complex fluids is extremely limited. Understanding how low temperatures and segregation from ice-rich domains impacts the structural stability of soft self-assembled systems could be instrumental in domains as diverse and as relevant as marine biology, cryobiology or food science, among others.[10] Whether harsh freezing conditions disrupt soft self-assembled fluids or whether the mechanical stability after casting depends on the spatial organization of the fluid before freezing is an open question.

Despite being historically associated with applied materials, freeze-casting has recently enabled to tackle more fundamental physicochemical questions in soft matter such as the study of micellar phase transitions at sub-ambient temperatures, a regime often overlooked at. Albouy *et al.*[11] have unveiled the phase diagram of P123 block copolymer solutions at low temperature using ice-templating, effectively showing that phase transitions are possible in the interstitial space between ice crystals. Combining freeze-casting with classical soft compounds such as phospholipids (DMPA)[12] or block copolymers (P123)[13] in the presence of soluble silica precursors leads to silica materials, displaying hierarchical porosity and good stability due to the mineral content.

Supramolecular hydrogels composed of low molecular weight gelators constitute an important class of materials for their potential applications in domains as diverse as food science, biomedicine and tissue engineering.[1,14–18] The main interest of these soft materials is constituted by their self-assembled nature, making the gelation fast and reversible. However, the main drawback is constituted by their sensitivity to physicochemical parameters controlling the self-assembly, and the window of practical interest is often very limited and it strongly depends on the type of molecule employed.[18] It goes without saying that not any material processing is adapted to these conditions[19] and freeze-casting is probably one of the least obvious process that can satisfactorily enhance the sensitive mechanical stability of self-assembled hydrogels. This probably explains the reason why, to the best of our knowledge, ice-templating at very low temperatures (e.g., < -20°C) was never used to cast self-assembled hydrogels into soft macroporous fibrillar foams. The only example of a frozen, although uncast, self-assembled fibrillar network (SAFiN) composed of the low molecular weight gelator (LMWG) Fmoc–Phe–Phe was shown to withstand a temperature of -12°C without apparent disassembly of the fibers,[20] even if the issue of the mechanical stability after drying was not addressed.

In this work, we show that freeze-casting has a dramatic impact on the mechanical properties of an isotropic dispersion of entangled self-assembled fibers, constituting a hydrogel at room temperature. On the contrary, freeze-casting improves the mechanical properties of a kinetically-trapped lamellar hydrogel. Both self-assembled networks are constituted by pH-responsive bio-based biocompatible glycolipids of similar chemical structure and their respective hydrogels are both prepared in water by a pH-jump method. Additionally, both hydrogels are constituted by flat structures (twisted ribbons against interdigitated bilayers) and the only difference is the origin of the constraint in the material at room temperature: entanglement for the fibrillar hydrogel and structural defects in the lamellar stacking.

In summary, this work highlights how the mechanical stability of two macroporous foams obtained from self-assembled chemically-analogous hydrogels strongly depends on the nanoscale structure of the gels and in particular on the physical origin of the mechanical constraints. In particular, the rare class of lamellar hydrogels can provide unprecedented soft spongy solid foams with unexpected isotropic mechanical behavior and Young moduli in the same order as those measured in some living tissues.

**Results and Discussion**

*Self-assembled fibrillar and lamellar hydrogels can be cast into 3D foams*. The self-assembly properties of SLC18:0 and GC18:0 glycolipids under dilute conditions have been extensively described in previous works:[21–23] at concentration values below 1 wt%, room temperature and pH 6.2 ± 0.3. SLC18:0 forms flat twisted nanoscale ribbons having a side cross-section of about 13 nm and "infinitely" long (compared to the cross-section) in the longitudinal dimension (SLC18:0 in Figure 1). GC18:0 forms "infinitely" extended (compared to the thickness) flat interdigitated layers (IL) with a thickness of about 3 nm (GC18:0 in Figure 1). At higher concentrations both samples form hydrogels.[24,25] Interestingly, the bulk mechanical properties are comparable for both systems. The frequency-dependent elastic ($G'$) and loss ($G''$) moduli measured for both hydrogels at C= 5 wt% (Figure 1) never cross, with $G'>G''$ (typical of a gel) in the entire frequency range. The order of magnitude of $G'$ is comparable for both hydrogels, being contained between 0.3 kPa and 1 kPa in the high frequency regime and equalizing below 0.1 rad.s$^{-1}$.

Despite the comparable values of the elastic moduli, the frequency-dependent evolution of the storage and loss moduli is different, thus reflecting the strong structural diversity between the two materials. SAXS[22,24,26] and cryo-TEM[22] (Figure 1a) experiments show the fibrillar crystalline and isotropic nature of SLC18:0 samples, of which the rheological behavior at 5 wt% shows that $G'(\omega) > G''(\omega)$, with no evidence of angular frequency dependence of the storage modulus $G' \propto \omega^0$, indicating that SLC18:0 forms gels over the entire angular frequency range (Figure 1c, red diamonds). When plotting the plateau elastic modulus, $G_0$, as a function of concentration $C$, $G_0 \propto AC^n$ ($A$ being a constant and $n$ an empirical exponent), a well-known scaling law measured in colloidal and polymer gels,[27–29] one finds values for $n$ between 2.0 and 2.4,[24] indicating that the stability of SLC18:0 is governed by fiber entanglement. Similar empirical $G_0(C)$ behavior is in a good agreement with scaling laws found for isotropic fibrillary hydrogels composed of bacterial cellulose[30] and low molecular weigth gelators,[30–33] thus describing the structure of SLC18:0 as a typical isotropic fibrillar hydrogel.

As far as the GC18:0 sample is concerned, SAXS[23,25,26] and cryo-TEM[26] (Figure 1b) demonstrate the presence of flat lamellar structures. Upon increasing the concentration, GC18:0 also forms hydrogels, of which the typical viscoelastic response follows a power-law behavior over four orders of angular frequency magnitude (Figure 1c). Both moduli scale as $G' \propto G'_0 \, \omega^\alpha$, and $G'' \propto G''_0 \, \omega^\beta$, where $G'_0$ and $G''_0$ are pre-factors at $\omega = 1$ rad·s$^{-1}$ and $\alpha$ and $\beta$ the exponents, between $0.15 < \alpha, \beta < 0.3$, for mechanically stable GC18:0 hydrogels. Such distinctive power-law rheology response is not uncommon for complex materials, and is generally described using fractional rather than canonical rheological models.[34] It indicates a broad distribution of relaxation times,[35] modeled for critical gels[36] and soft glassy materials.[37] The latter in particular considers that viscoelasticity is controlled by disorder, metastability and local structural rearrangements between the mesoscopic elements.[38] Thermal motion alone is not sufficient to reach complete relaxation and the system has to cross energy barriers, larger than typical thermal energies.[38] In the present case, optical and confocal laser scanning microscopy performed on GC18:0 hydrogels at 5 wt% and 2.5 wt% highlighted a complex structure characterized of multiscale lamellar domains (of size between 100 µm and 500 µm and below 5 µm), coexisting with spheroids containing disordered lamellar sheets.[25] In this context, and even if a direct relationship between microstructure and mechanical response remains challenging,[39] the power-law behaviour of $G'(\omega)$ and $G''(\omega)$ most likely describes a heterogeneous environment characterized by the rearrangement of the lamellar domains and a very broad range of microstructural length constituting GC18:0 hydrogel.

Figure 1c summarizes then the elastic properties and bulk structure of both SLC18:0 fibrillar and GC18:0 lamellar hydrogels prior to freeze-casting: the bulk mechanical properties are comparable ($G'$ ~ 0.3-1 kPa range) at pH 6.2 ± 0.3 but the local structure is extremely different: the fibrillar hydrogel can be described by a homogeneous dispersion of entangled fibers, while the lamellar hydrogel is best described by a heterogeneous dispersion of interconnected lamellar domains. A more comprehensive description of the structural and mechanical properties of SLC18:0 fibrillar and GC18:0 lamellar hydrogels is reported elsewhere.[24,25]

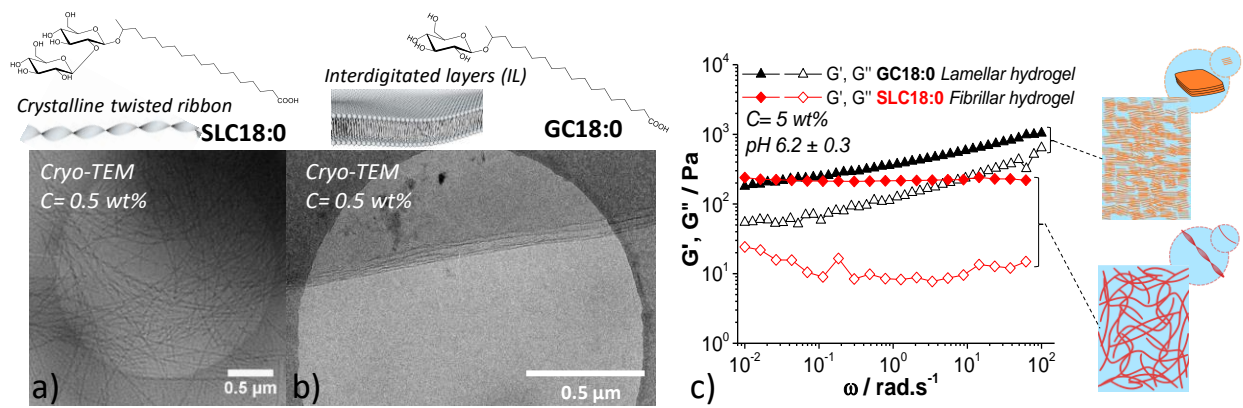

**Figure 1 – a-b) Cryo-TEM images of the typical a) fibrillar and b) lamellar structures of SLC18:0 and GC18:0 observed under diluted conditions (C= 0.5 wt%) and pH 6.2 ± 0.3. c) Angular frequency, $\omega$, dependence of elastic, $G'$ (closed symbols), and loss, $G''$ (open symbols), moduli for fibrillar SLC18:0 (red diamonds) and lamellar GC18:0 (black triangles) hydrogels recorded in the linear regime ($\gamma$ = 0.02% and 0.05%, respectively) at pH 6.2 ± 0.3 and C= 5 wt% prior to freeze-casting. The drawings show the respective bulk structure of the hydrogels: homogeneously entangled fibers for SLC18:0 and heterogeneous dispersion of interconnected large lamellar domains for GC18:0.**

Despite the nanoscale structural difference between them, freeze-casting[1] of both fibrillar SLC18:0 and lamellar GC18:0 hydrogels result in the formation of macroporous solids, as shown by the corresponding SEM-FEG images in Figure 2. The longitudinal ∥$Z$[1] (Figure 2 - L1-L3) and transverse ⊥$Z$[1] (Figure 2 - T1-T3) sections of a typical freeze-cast GC18:0 hydrogel show an intertwined, spongious, network with thick (order of the micrometer) walls. ∥$Z$ section at low magnification (Figure 2 - L1) shows the classical alignment in the direction of ice growth, while ⊥$Z$ section at equivalent magnification (Figure 2 - T1) shows a spongy structure. However, a closer look at higher magnifications both in the ∥$Z$ (Figure 2 – L2-L3) and ⊥$Z$ (Figure 2 – T2-T3) sections displays a more homogeneous cellular 3D network. The presence of a pseudo isotropic 3D spongy network in GC18:0 foams (marked with a fluorescent Lissamine B-grafted lipid) is confirmed by confocal laser scanning microscopy (CLSM) experiments performed on sections ∥$Z$. Figure 2a,b and Video 1 and Video 2 provided as Supporting Information distinctively show a homogeneous interconnected network, where the direction of freezing (∥$Z$) can hardly be identified. From CLSM data one can extract the orientational distribution of the intensity, $I(\varphi)$, and the corresponding degree of anisotropy, defined as $\frac{I(\|Z\equiv 0°)}{I(\perp Z\equiv 90°)}$; for the GC18:0 foams $\frac{I(0)}{I(90)} < 4$ (Figure S 1). This is an unexpected behaviour, if compared to more classical freeze-cast polymer-based systems,[40–43] which are characterized by a well-defined anisotropic arrangement of matter between the longitudinal ∥$Z$ (walls) and transverse ⊥$Z$ (connected pores) directions. On the contrary, the spongy network of GC18:0 seems less anisotropic and more homogeneous: the discrepancy between the 3D organization of matter is much less pronounced between ∥$Z$ (walls) and ⊥$Z$ (pores) sections. In particular, L2, T2 and L3, T3 panels in Figure 2 show a homogeneous distribution of open pores and walls both in the ∥$Z$ and ⊥$Z$ directions.

---

[1] The direction of the ice-growing front in the freeze-casting device is identified as the Z-axis throughout the paper (Figure S 3). Nomenclatures ∥$Z$ and ⊥$Z$ respectively refer to longitudinal and transversal planes of observation with respect to the ice-growing front in the Z-axis.

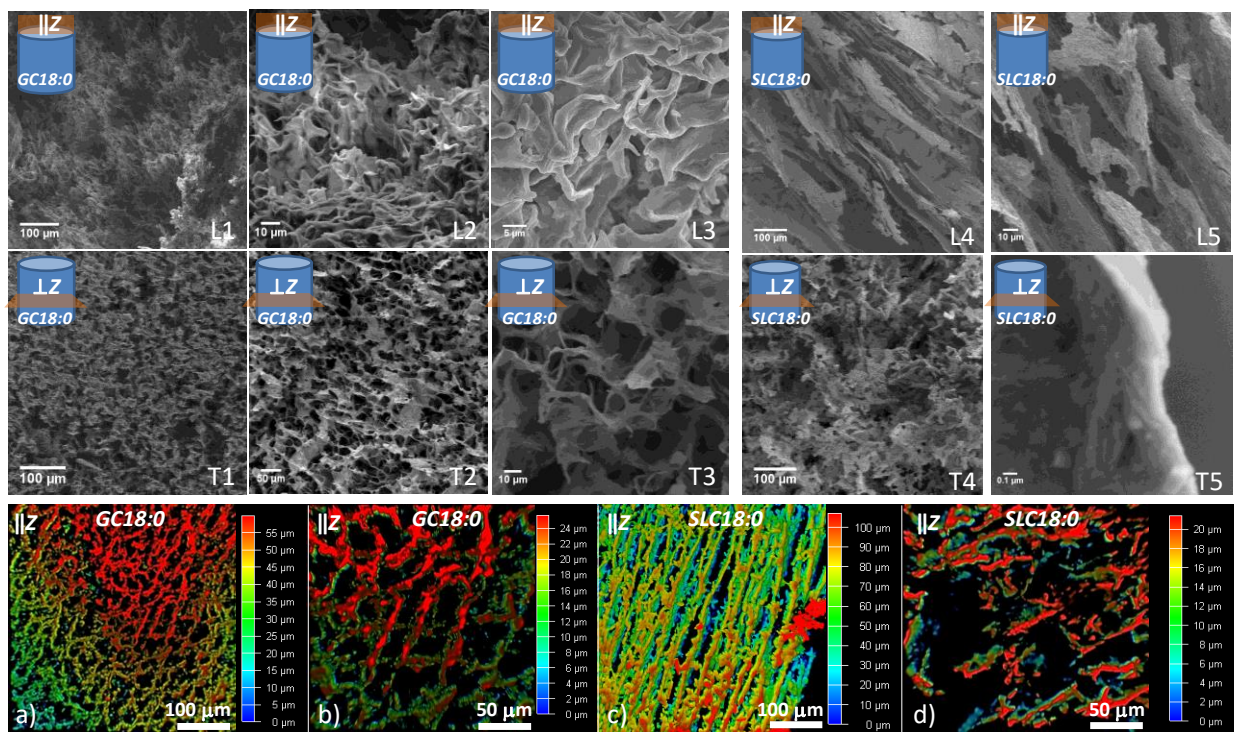

**Figure 2** – SEM-FEG images of longitudinal ∥$Z^1$ (L1-L3) and transverse ⊥$Z^1$ (T1-T3) sections of GC18:0 solid foams prepared at a freezing rate of 5°C.min$^{-1}$. All samples are prepared at C= 10 wt% except in panels T2 and T3, where concentration is C= 5 wt%. SEM-FEG images of ∥Z (L4-L5) and ⊥Z (T4-T5) sections of SL18:0 solid foams prepared at C= 10 wt% and freezing rate of 5°C.min$^{-1}$. All samples were in a hydrogel state before freeze-casting and drying. a)-d) 3D images displaying typical confocal laser scanning microscopy (CLSM) experiments performed on ∥Z sections of GC18:0 and SLC18:0 (C= 5 wt%) solid foams (freezing rate: 5°C.min$^{-1}$). Additional CLSM data (∥Z sections) for each class of material in the form of 3D videos are shown as Supporting Information.

The corresponding longitudinal ∥Z and transverse ⊥Z sections of the macroporous solids obtained from SLC18:0 hydrogels (C= 10 wt%, pH 6.2 ± 0.3) are presented in Figure 2 – L4-L5, T4-T5. ∥Z sections (Figure 2 – L4-L5) show the presence of extended (hundreds of microns) portions of walls, stacked onto each other, poorly interconnected and massively ∥Z. ⊥Z (Figure 2 – T4), on the contrary, is characterized by an open macroscale porosity. The thickness of each wall can be estimated in the range of the micrometer (Figure 2 – T5). CLSM experiments performed ∥Z (Figure 2c,d, Video 3 and Video 4 as Supporting Information) confirm that these foams are composed of highly anisotropic open structures, characterized by parallel flat walls of about 1 μm to 2 μm in thickness and separated by a void longitudinal space, of size between 25 μm and 50 μm. The 3D analysis done by CLSM ∥Z over several hundred micrometers (Video 3 and Video 4 as Supporting Information) show that the transversal connection between the walls is poor, if non-existing in some portions of the material. The orientational distribution is much more pronounced, with $\frac{I(0)}{I(90)} > 10$, that is more than twice as found in lamellar foams (Figure S 1). Overall, SLC18:0 solids have a preferential orientation of the macropores ∥Z, as expected and experimentally-found in practically all freeze-cast soft and inorganic matrices.[5,40,41]

In summary, microscopy highlights the strong morphological difference at the micrometer scale between the freeze-cast SLC18:0 macroporous solid and GC18:0 solid foams: the former are composed of stacked fibrillar flat sheets (of thickness in the micrometer range) ∥Z, while the latter are composed of a highly interconnected, disordered, sponge-like 3D network, poorly aligned ∥Z.

*Fibrillar and lamellar nanostructures are not affected by freeze-casting above ~100 nm and below ~10 nm*. The possible disruption of the nanoscale fibrillar and lamellar assemblies of the glycolipids by the harsh freeze-casting conditions is a crucial issue in view of developing macroporous soft materials, and this issue will be addressed hereafter at scales above ~100 nm and below ~10 nm. The literature is not abundant on this matter, but it was recently shown that ice crystallization induces phase transitions in block copolymer micellar systems,[11] that is it pushes the initial thermodynamic phase out of equilibrium, provoking molecular rearrangement. In the systems presented here, one can expect that rapid temperature variations combined with immediate compression of matter could reasonably push the SLC18:0 and GC18:0 out of their respective fibrillar and lamellar phase regions towards an unknown (at the present state of the art for these molecules) phase. In addition, anisotropic ice crystallization is equivalent to a dehydration process in the vicinity of the solute, inducing concentration changes of both salt and hydronium ions: the former process is known to have strong impacts on the elastic properties of lamellar phases,[44–48] while the latter could induce a raise in the local pH and hydrolysis of the glycosidic bonds. Both processes are generally described at room temperature, but it is unclear whether they could occur at temperatures much below zero.

FT-IR (Figure S 2), recorded for both SLC18:0 macroporous solid and GC18:0 solid foams to verify the molecular integrity, presents the typical spectral features of these compounds,[49] and in particular, the CH symmetric and antisymmetric stretching bands ($\nu= 2927$ cm$^{-1}$, 2860 cm$^{-1}$), the C=O stretching both in the COOH ($\nu= 1730$ cm$^{-1}$), the glucosidic C–O–C,

C–O stretching, coupled with C–C stretching and O–H deformation ($\nu$=1162 cm$^{-1}$, 1077 cm$^{-1}$, 1027 cm$^{-1}$)[50] and the region below 1000 cm$^{-1}$, generally attributed to the carbohydrate ring vibration and breathing mode ($\nu$= 935 cm$^{-1}$, 729 cm$^{-1}$) and anomeric carbon deformation ($\nu$= 898 cm$^{-1}$).[50] The attribution of the peaks in the 1600 cm$^{-1}$ – 1250 cm$^{-1}$ region should be more careful; for instance, the peaks at 1565 cm$^{-1}$, 1416 cm$^{-1}$ and 1386 cm$^{-1}$ could be attributed to the asymmetric and symmetric stretch of the carboxylate anion,[51] which can reasonably be still present as a residue at pH around 6. However, if one excludes the presence of residual COO$^-$ groups, CH bending (1380 cm$^{-1}$) and twisting (1268 cm$^{-1}$) and OH in-plane bending (1400-1455 cm$^{-1}$) could also be possible attribution.[50,52] All in all, the data obtained by FT-IR shows that the controlled freezing process does not degrade the glycolipids.

The molecular (< 1 nm) arrangement and local order (< 10 nm) of the glycolipids is probed by a combination of XRD and SAXS. The XRD profiles (please note that the XRD scale is voluntarily reported in nm$^{-1}$, for a matter of convenience with SAXS data) of SLC18:0 macroporous solids (Figure 3d) show a strong diffraction peak at $q$ = 2.28 nm$^{-1}$, corresponding to a $d$-spacing of 2.75 nm $d = 2\pi/q$), commonly attributed to the repeating distance between SLC18:0 molecules within the plane of a flat twisted fiber.[22] This peak is only 0.2 nm$^{-1}$ off (8%) with respect to the typical values measured on the same sample in solution. The small difference of only 0.24 nm between the macroporous solid and solution sample, if not negligible, could be attributed to a small variation in the tilting angle of the SLC18:0 molecules within the fiber plane.[22] One should additionally observe that the freezing rate has no influence on the peak position. The fibrillar twisted structure of the SLC18:0 macroporous solids is nicely confirmed at scales above 100 nm by high magnification SEM-FEG images recorded on two samples prepared at freezing rates of 5°C.min$^{-1}$ (Figure 3e) and 1°C.min$^{-1}$ (Figure 3f). The surface characteristics of the SLC18:0 macroporous solids demonstrate the presence of a fibrous texture (surface mean squared roughness, R$_{RMS}$, of 20.3 and 24.4 for, respectively, the 5°C.min$^{-1}$ and 1°C.min$^{-1}$ samples), where regular pitches, separated by about 50 nm and 100 nm, can distinctively be identified along the fibers' longest axes, in agreement with the twisted ribbon structure prior to freezing (Figure 1).[22] An estimation of the fiber cross-section gives values between 15 nm and 25 nm, which also characterize this sample in solution (Figure 1).[22] SEM-FEG shows that fibers do not aggregate into bundles but they are rather entangled within the walls of the solid foam.

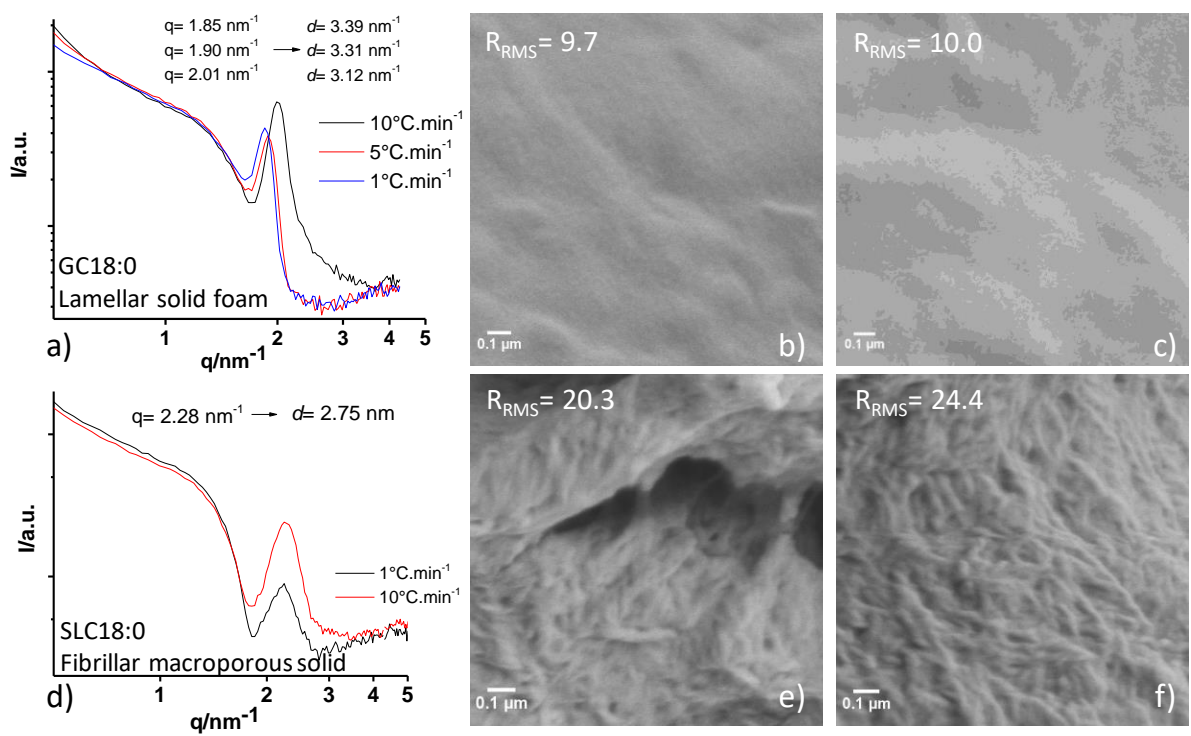

**Figure 3** – a,d) X-ray diffraction (θ-2θ geometry; λ= 1.54 Å; 2θ has been converted into a q/nm$^{-1}$) patterns of (a) GC18:0 solid foams and (d) SLC18:0 macroporous solids (both at C= 10 wt%), freeze-cast at various freezing rates. b,c-e,f) High magnification (x50000) SEM-FEG images of GC18:0 solid foams (b,c) and SLC18:0 macroporous solids (e,f) prepared at C= 5 wt%. Freezing rate is (a,c) 5°C.min$^{-1}$ and (b,d) 1°C.min$^{-1}$. All samples were in a hydrogel state before freeze-casting.

A closer look at the nanoscale structure of GC18:0 solid foams show much smaller $R_{RMS}$ (Figure 3b,c), in the order of 10, that is about 2.5 smaller than in the case of SLC18:0 foams, indicating a much smoother surface, with wavy texture, as shown by SEM-FEG images. This is in agreement with a material composed of flat bilayers, as observed by cryo-TEM prior to freeze-casting (Figure 1).[23] The typical XRD patterns of the solid foams display a dependence between lamellar *d*-spacing ranging between 3.1 nm and 3.4 nm, calculated in the q-region between 1.8 nm$^{-1}$ and 2.0 nm$^{-1}$, and the freezing rates (Figure 3a): the faster the freezing rate, the smaller the *d*-spacing. The typical SAXS fingerprint of GC18:0 bilayers in solution[23,26] does not show any diffraction peak in this q-range, which is rather characteristic of a broad oscillation due to the form factor of the bilayer. We have shown that GC18:0 lamellar hydrogels display broad (100) and (200) reflections at much lower *q*-values,[25] corresponding to *d*-spacings between 15 nm and 25 nm. These large periods are typically observed in lamellar phases stabilized by long-range repulsive steric and/or electrostatic interactions.[53–56] It is well-known that *d*-spacing in a thermodynamic lamellar phase strongly fluctuates according to a broad and complex number of parameters including osmotic stress regulating the bilayer hydration, bilayer charge density, salt concentration, temperature.[57–61] In light of this, we make the hypothesis that the small *d*-spacings found in the solid foams are highly sensitive to freezing, that is, to dehydration of the interbilayer water during freeze-casting, and, consequently, to the freezing rate. This hypothesis was tested by performing a comprehensive study through temperature-resolved *in-situ* small-angle X-ray scattering (SAXS) experiments (Figure 4), adapting a specifically-conceived freeze-casting cell to the SAXS beamline ID02 at the ESRF (Grenoble, France) (for more information, please refer to Figure S 3).

Temperature-resolved *in-situ* SAXS experiments recorded for the GC18:0 (10 wt%) at 10°C/min (contour plot, Figure 4a) show the coexistence of the (100) peak (intense red signal) at $q = 0.34$ nm$^{-1}$ ($d= 18.5$ nm) and the oscillation of the bilayer form factor (*ff*; broad green signal) between 2 nm$^{-1}$ and 4 nm$^{-1}$ in the temperature range between 20°C and ~ -8°C, in agreement with previous experiments at room temperature.[62] Below -10°C, the (100) reflection undergoes a sharp transition from $q= 0.37$ nm$^{-1}$ ($d= 17.0$ nm) to $q= 1.32$ nm$^{-1}$ ($d= 4.8$ nm, intense green signal), thus masking the form factor of the bilayer. This trend is systematic for any freezing rate and position in the freezing cell, as shown by the patterns presented in Figure 4b (freezing rate of 5°C/min) for selected temperature values. Figure 4a shows that: a) the temperature-driven transition of the diffraction peak between low-*q* and high-*q* regions is very fast, as also confirmed by the evolution of *d*-spacing values (Figure 4c) for various freezing rates; b) three distinct regions can be identified, a *low-q (1)* and *high-q (3) regime*, respectively reflecting the lamellar order above 15 nm and below 5 nm, but also a *transition regime (2)*, depicting the transition between (1) and (3). To better explore the latter, which is contained within less than three degrees, we have run an experiment (Figure 4b1) employing a faster data acquisition rate, 150 frames.°C$^{-1}$, instead of the standard 0.2 frames.°C$^{-1}$ used for all *in-situ* SAXS experiments performed in this work. Figure 4b1 shows two regimes having low-*q (1)* and high-*q (3)* lamellar scattering, respectively above -9°C and below -11°C, as discussed previously; meanwhile, for temperature values within this range, which we defined as *transition regime (2)*, there is neither any clearly observable diffraction profile nor any structural continuity between the *low-q* and *high-q regimes*, as one could expect. A topological interpretation of the *transition regime (2)* will be given below after commenting the 2D plot profiles.

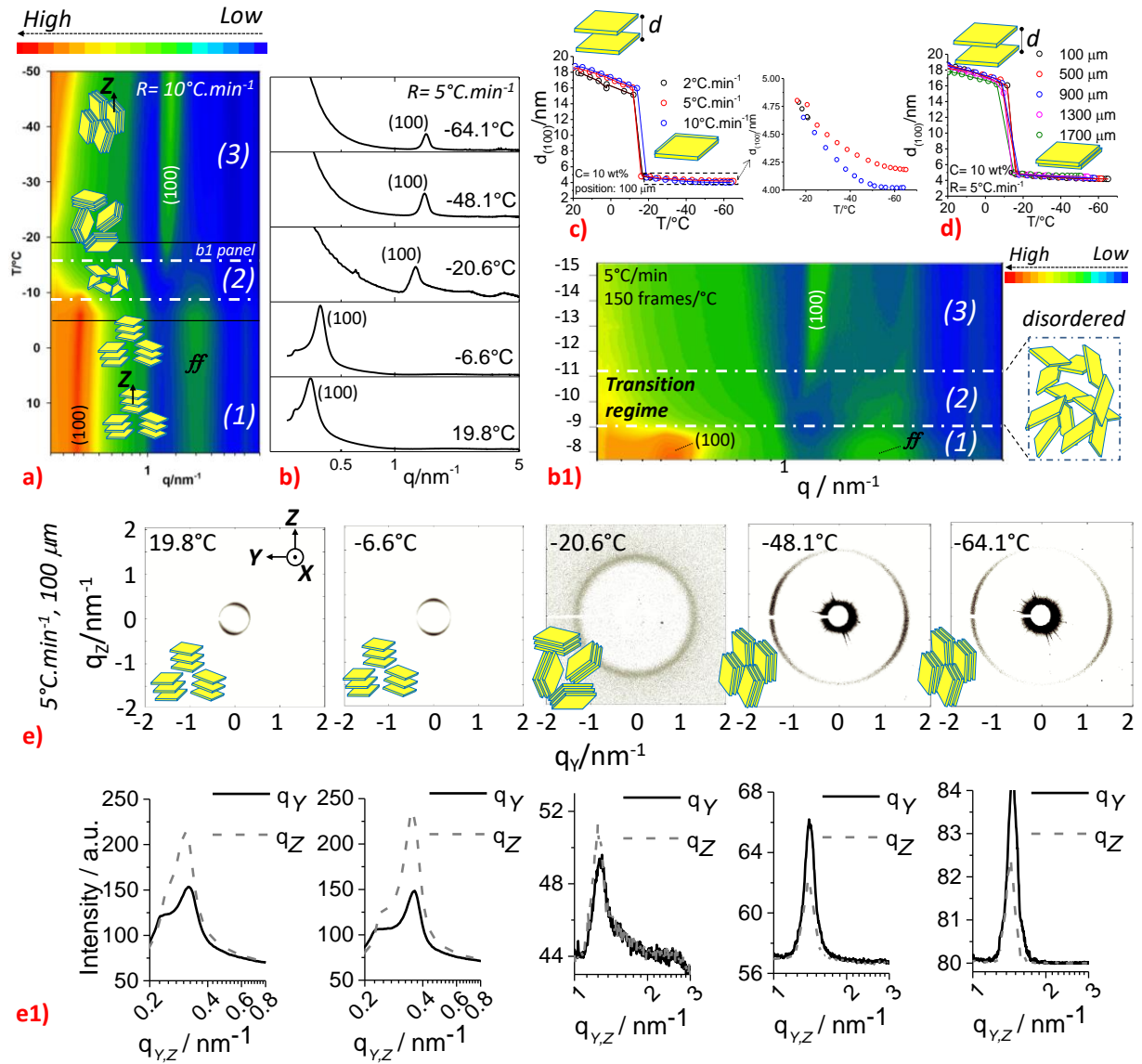

**Figure 4 – a,b)** Temperature-resolved *in-situ* SAXS experiments of a GC18:0 hydrogel (C= 10 wt%) undergoing freeze-casting between T= 20°C and T= -65°C at a freezing rate of a) 10°C.min$^{-1}$ (contour plot) and b) 5°C.min$^{-1}$ (selected patterns), using a data acquisition rate of 0.2 frames.°C$^{-1}$. **The lamellar peak is identified with the classical (100) Miller index, while the *ff* notation identifies the oscillation of the lamellar form factor.** b1) *In-situ* SAXS experiment performed in b) and repeated with a faster data acquisition rate of 150 frames.°C$^{-1}$. *(1), (2)* and *(3)* respectively identify the *low-q, transition* and *high-q* regimes. **c-e)** Evolution of the *d*-spacing as a function of freezing rate (position: 100 µm from the metal bar, Figure S 3b) and position from metal bar in the freeze-casting device in a GC18:0 hydrogel (C= 10 wt%). The 2D SAXS patterns are extracted at specific temperatures from the sample cooled at a freezing rate of 5°C.min$^{-1}$ at the position of 100 µm from the metal bar. Orientation of the lamellae is indicated by the cartoon within each 2D pattern. The direction of the ice-growth occurs along the Z-axis. e1) $I(q_Z)$ and $I(q_Y)$ intensity distribution profiles extracted for each 2D plot in e). A specific freeze-casting cell has been conceived and adapted to the SAXS beamline (ID02, ESRF, Grenoble, France) to run this experiment. Description of the cell and images of the setup are given in Figure S 3. One should refer to Figure S 4 for the procedure to recover the values of *d*-spacing as a function of temperature at a given position.

Figure 4c and Figure 4d show the evolution of the lamellar *d*-spacing in the GC18:0 lamellar hydrogel (10 wt%) at room temperature during the freeze-casting process, as a function of the freezing rate (at a position of 100 µm) and position (freezing rate settled at 5°C/min) in the freeze-casting cell. The equilibrium *d*-spacing at 20°C is generally contained between 18 nm and 20 nm, as found in previous works,[23,25,26] and it decreases to about 16 nm close to the freezing temperature, at about -15°C. After freezing, *d*-spacing drops to about 4.8 nm and it slowly decreases to about 4.0 nm at -60°C. It is interesting to note that, if the water-ice transition temperature, identified where *d*-spacing undergoes a sharp drop, is neither sensitive to the freezing rate nor to the position in the cell, the actual values of the *d*-spacing are sensitive to the freezing rate, as shown by the inset in Figure 4c: the faster the freezing rate, the smaller the *d*-spacing. This is in good agreement with the diffraction data presented in Figure 3a. On the contrary, the *d*-spacing is absolutely independent on the position in the freezing cell (Figure 4d). These data confirm the analogy between the effect of freeze-casting on the *d*-spacing in the GC18:0 lamellar phase with very well-known effects of dehydration and osmotic stress in thermodynamic lamellar systems.[53,54,56,63,64]

Finally, the 2D SAXS profiles (5°C/min, 100 µm, Figure 4e), and the corresponding $I(q_Y)$, $I(q_Z)$ (Figure 4e1) and azimuthal $I(\varphi)$ (Figure S 5) intensity distribution profiles, show the orientation of the lamellar hydrogel in the cell. In the reference system (drawn both in Figure S 3c and Figure 4e) the incident x-ray beam is ∥X while freezing occurs ∥Z. The orientation of the lamellae in this frame is drawn for each temperature in Figure 4e. From T= 19.8°C to -6.6°C, the scattering is strongly accentuated ∥Z, meaning that the lamellae are contained in the corresponding transverse plane of the incident x-ray beam (⊥Z). After freezing, at T= -20.6°C, the diffraction ring is isotropic: the long-range correlation is kept but anisotropy is lost. At lower temperatures (T= -48.1°C, -64.1°C) the diffraction is reinforced ∥Y, indicating that lamellar phase has tilted of a 90° angle: the lipid layers are now contained in the *XZ* plane (⊥*Y*). Although not systematic, we observe such a reorientation of the lamellar phase in most of our experiments, and it indicates that upon freezing, the preferential orientation of the lamellar phase is tilted by the ice front. These considerations are also supported by the evolution of the orientation order parameter S for the different q values and temperatures (Figure S 5c).

Two questions arise: through which topological mechanism reorientation occurs and what happens to the lamellar order in the *transition regime* (-9 < $\Delta$T/°C < -11 in Figure 4e1)? Two hypotheses can be formulated, one involving continuity and the other discontinuity. In the first one, the ice front tilts the bilayers in the *YZ* plane, transverse to the beam direction, before tilting them in the *XZ* plane at lower temperatures. During this process, one does not expect to observe any SAXS signal when the lamellae are in the *YZ* plane. However, this hypothesis is not satisfying, because it cannot explain the isotropic order below -11°C: if the ice front simply tilts the lamellae, one would expect an anisotropic signal in the *XZ* plane immediately after freezing, which is not the case. In the second hypothesis, ice formation disrupts the lamellar organization and long-range order is lost for a short length of time, after which the lamellar phase undergoes topological reorganization. This hypothesis is more realistic, because it could explain the isotropic scattering signal after freezing. The water-ice phase transition withdraws water from the interlamellar layer (*d*-spacing> 15 nm before freezing); the bilayers, instead of undergoing an immediate collapse, pass through a disordered phase. When water has frozen, and the bilayers have dehydrated, the system is composed of randomly-oriented collapsed (*d*-spacing < 5 nm) lamellar domains, providing the isotropic signal, analogous to a powder diffraction. Despite the lowering temperature from -11°C to -60°C and the confinement in between the ice crystals, the lamellar domains remain fluid enough to slowly reorient themselves in the *XZ* plane, probably due to a small fraction of unfrozen water. The cartoons superposed to Figure 4a and Figure 4b1 help the reader visualizing these effects.

In summary, the combination of SEM-FEG and XRD/SAXS, respectively exploring size domains > 100 nm and < 10 nm, show that freeze-casting does not affect the nature of the fibrillar and lamellar assembly observed in the hydrogels at room temperature prior to freezing, but it simply compacts them into a more confined space contained within the ice domains. At the moment, we have no evidence of any phase transition induced by temperature and the process is entirely reversible: we have reversibly freeze-cast and heated the same sample several times and we have never experienced appreciable differences in terms of *d*-spacings before and after freezing. For the SLC18:0 macroporous solids, freezing has no practical influence on the molecular arrangement of SLC18:0, but simply on the fibrillar confinement in a much narrower volume. This is easily explained by the difference in term of self-assembly between GC18:0 and SLC18:0. The former forms a lamellar phase composed of interdigitated layers, where GC18:0 molecules are settled orthogonally to the bilayer plane.[23,26] The bilayers are separated by water and their distance is controlled by electrostatic forces,[25] very sensitive to hydration and salt concentration. In the latter, twisted ribbons are rather described as semicristalline fibers, into which SLC18:0 molecules are assembled in the fiber plane.[22] This system is much less sensitive to dehydration due to its semicrystalline state.

*Fibrillar macroporous solids are brittle; lamellar solid foams are spongy and standing up to 1000 times their own weight.* If freeze-casting does not affect the fibrillar and lamellar morphologies, one could reasonably expect that the mechanical properties of the corresponding fibrillar macroporous solids and lamellar solid foams are comparable. However, the mechanical tenure of the two types of materials is very different: all SLC18:0 macroporous solids prepared in this work are extremely sensitive to delamination ∥Z, making them brittle and fragile. On the contrary, all GC18:0 foams are rather spongy and they can be easily manipulated. The first striking result is that no mechanical measurement could be performed on the SLC18:0 macroporous solids below 5 wt% due to their brittleness: the simple recovery of the sample from the polypropylene tube applies a stress, large enough to break the sample apart. At concentrations above or equal to 5 wt%, the sample is more resistant, as qualitatively shown in Figure 5c, and stress-strain compression experiments performed ∥Z (Figure 5b) can display Young moduli between 5 kPa and 15 kPa, for concentrations between 5 wt% and 10 wt%. However, compression ⊥Z (Figure 5b) provide a Young's modulus below 0.5 kPa. Interestingly, the stress-strain profile ⊥Z (Figure S 6b) is oscillatory, suggesting a series of regular fractures at the micrometer scale, compatible with the anisotropic planar structures observed in Figure 2c,d and Video 3,4. A control experiment, performed on a SLC18:0 fibrillar hydrogel frozen in an isotropic environment (freezer at -80°C), shows comparable stress-strain curves measured ∥Z and ⊥Z, with moduli below 0.5 kPa (Figure S 6c). Such strong mechanical anisotropy after the freeze-casting process is expected, as shown for pectin foams.[40]

Stress-strain experiments were also performed on lamellar solid foams both ∥Z and ⊥Z (Figure 5a,b). Typical compression experiments ∥Z show a behavior with two distinctive regimes: a linear Hookean regime followed by a densification regime, where the stress increased significantly. As expected, the densification regime occurred at lower strains as function of glycolipid amount (~85%, ~70% and ~60% for 1, 5 and 10 wt%, respectively). Due to the low compression speed, no evidence of load failure could be detected from the different curves. The inset graph in Figure 5a represents the magnification of the data in the small strain range (≤ 5 %) and showed that the external stress increased as function of the concentration of GC18:0. The Young's moduli of the solid foams measured ∥Z are also found in the range of several kPa and we could measure values as high as 30 kPa (Figure 5b) at 5 wt%. However, if the dispersion of Young's moduli is quite high, probably due to the strong variability in the foam homogeneity from one experiment to another, compression experiments performed ⊥Z generally display a comparable slope to ∥Z (Figure 5b), suggesting an isotropic distribution of the Young's moduli in the volume of the GC18:0 foams. A control experiment, performed on a GC18:0 lamellar hydrogel frozen in a isotropic environment (freezer at -80°C), also show stress-strain curves of comparable slope when measured ∥Z and ⊥Z (Figure S 6a). Values of Young's moduli between 5 and 30 kPa are comparable with foams prepared by freeze-drying of decellularized adipose tissue,[65] and with breast tissues (fat and fibroglandular tissues) exhibiting Young's moduli of 3.25 kPa[66] and they reflect the soft nature of the solid foams, which behave as spongy solids at room temperature. However, the isotropic distribution of Young's moduli after freeze-casting is unexpected given the strong anisotropic processing conditions.

The striking difference in terms of mechanical anisotropy between solid foams obtain from fibrillar SLC18:0 (high anisotropy) and lamellar GC18:0 (low anisotropy) hydrogels are also qualitatively presented in Figure 5c: the interconnected 3D structure of the lamellar foam after drying can stand up to 1000x its own weight, while the fibrillar solid is crushed only between 50x and 100x its own weight. The approximate density of the lamellar foams in this study is 0.05 g.cm$^{-3}$ with the highest values of Young modulus being in the order of 30 kPa. Referring to classical Ashby diagrams,[67] generally established for cohesive materials, mechanical properties of self-assembled lamellar foams compare to those of soft polymer foams, including freeze-cast cross-linked chitin biopolymers.[9] However, one should be aware that a comparison of the mechanical properties of the lamellar foams presented here with materials in the literature is inappropriate at the moment, because Ashby diagrams do not include a class of materials solely composed of weak forces and, to the best of our knowledge, we are not aware of analogous lipid lamellar or fibrillary foams. The ability to elaborate self-supported macroporous materials showing elastic modulus in the kPa range (similar to brain tissue mechanical properties) may find further application in the elaboration of hMSC cell culture supports with neurogenic differentiation properties.[68,69] To this regard, we have performed a series of qualitative tests to evaluate the stability of a series of GC18:0 spongy solid foams prepared at 5 wt% (using 5°C/min during freeze-casting) after immersion in water. We find that a typical monolith of 1 x 1 cm (height x diameter) and aspect ratio (height/diameter) of 1 keeps an aspect ratio between 0.5 and 0.8 up to about 3 h under magnetic stirring and over 4 h using a more gentle mechanical plate stirrer in water. These data, which unfortunately we cannot compare with similar materials in the literature, show nonetheless the tight cohesion of the self-assembled foam after freezing and drying and its potential employment in aqueous media without losing their initial shape below an estimated time of 2 h.

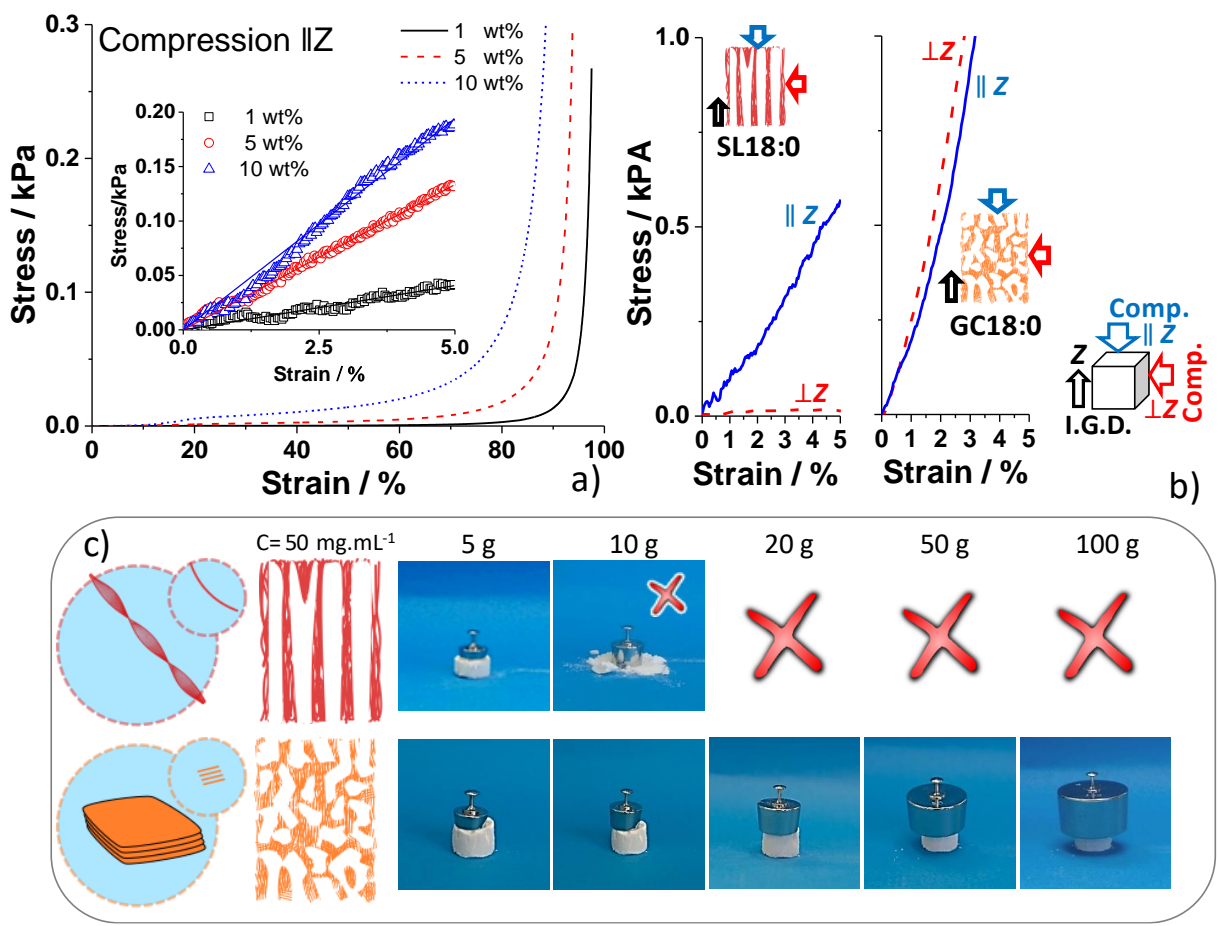

**Figure 5 -** Mechanical properties of the solid foams (ice has been removed). a) Stress–strain curves of GC18:0 lamellar foams at three different concentrations (1, 5 and 10 wt%) measured on cylindrical samples (Φ= 1.5 cm) by compression experiments ∥Z; the inset graph is the magnification of data used for estimating the Young's modulus (≤ 5% strain). b) Stress–strain curves of SLC18:0 fibrillar and GC18:0 lamellar foams at 5 wt% measured on cubic samples (1 cm x 1 cm x 1 cm) by compression experiments ∥Z (blue curves) and ⊥Z (red curves). Inset drawing: Comp.= Compression; I.G.D.= Ice Growth Direction ∥Z. c) Evidence of the superior mechanical properties of lamellar over fibrillar solid foams: the former can withstand up to 1000x their weight while the latter crushes at less than 100x its weight (here both foams have a weight of 100 mg).

*Discussion*. Before freeze-casting, fibrillar and lamellar hydrogels display comparable mechanical properties, with elastic moduli contained between 0.3 kPa and 1 kPa, but different microstructures: fibrillar gels are homogeneous and isotropic while lamellar gels are best described by a heterogeous structure held by interconnected lamellar domains. After the freeze-casting process, the mechanical properties depict a remarkable difference between the fibrillar and lamellar foams, the former being an exfoliating solid with poor resistance to compression ⊥Z and the latter bearing an isotropic response to compression. Interestingly, (fibrillar) isotropic and (lamellar) anisotropic gels respectively form foams with highly anisotropic and isotropic mechanical properties.

Self-assembled fibrillar network (SAFiN) hydrogels composed of low molecular weight gelators (LMWG) are generally accepted as isotropic network of entangled and/or partially branched fibers.[14,18,70] If the typical mechanical properties of SAFiN hydrogels are very well characterized in the temperature range between ~10°C and ~ 80°C, nothing is known of their mechanical properties at much lower temperatures, and in particular after freezing (and in particular freeze-casting) and drying. To the best of our knowledge, the only work, which processed SAFiN hydrogels at temperatures below zero (-12°C) did not specifically explore neither the orientational ice-templating nor the mechanical properties of the solid nor the stability of the fibers.[20] According to the SEM-FEG and CLSM analyses performed in the present work on the fibrillar solid foams (Figure 3 and Supporting Videos 3,4), the fibers are compressed into highly aligned walls, as expected.[5,10,40] If the mechanical properties of the hydrogel before freeze-casting rely on a classical isotropic network of entangled and branched fibers,[24] after freeze-casting the poor lateral cohesion between the wide fibrillar walls, separated by anisotropic voids of more than 25 nm in size, undoubtedly generates a large amount of dislocation defects along the ice-growth direction ($\parallel Z$). The latter are responsible for the exfoliation of the foams and their poor overall mechanical properties. Since XRD data indicate that the fibers crystal structure is not affected by the ice templating process, one could be tempted to make the hypothesis that the fibers are simply squeezed into micrometer-sized walls by the ice front during ice growth.

Unfortunately, the mechanical properties of freeze-cast (or simply frozen) self-assembled fibers are not discussed in the literature. However, one could compare our data to freeze-cast (or, frozen) nanocellulosic fibrillar materials. These systems show a number of analogies to SLC18:0 fibers: high aspect ratio, crystallinity, and glycosidic building blocks. However, all forms of nanocellulose materials (CNC, bacterial cellulose, etc…) have a significant difference with respect to SLC18:0 fibers: they are held by covalent glycosidic bonds within fiber plane. Foams prepared from nanocellulose are tough materials characterized by high compressive elastic moduli (above 100 kPa in some cases);[42,43,71–73] in-house control freezing experiments, performed on bacterial cellulose hydrogels (composed of crystalline nanofibers with similar diameter to the SLC18:0 ribbons) show the formation of hard foams, thus confirming the literature data. These data, corroborated by literature data performed on other anisotropic covalent materials, like carbon nanotubes,[74] show that ice crystallization cannot easily disentangle anisotropic covalent materials. In fact, it was recently shown that even colloidal crystals can form elastic foams, provided an embedding cross-linked covalent polymeric matrix, in the absence of which, the material behaves as a plastic monolith undergoing mechanical failure under modest strain.[75]

Obviously excluding cross linking, the action of ice crystallization could have two main effects on SLC18:0 hydrogels, namely disentanglement and/or fiber breaking. If disentanglement is not excluded, this phenomenon alone cannot explain the fragility of SLC18:0 fibrillar foams, because, similarly, one should also observe the formation of fragile nanocellulosic (or nanotube) foams, which is not the case, according to the literature survey. By excluding disentanglement, the only hypothesis that can explain the formation of poorly interconnected fibrillar walls must assume two combined effects at two different length scales: intra-fibrillar failure during freezing (from 10 nm to ~100 nm) and stiff walls (> 1 μm). High pressures during freezing could be enough to overwhelm the weak interactions (hydrophobic and H-bonding) keeping the SLC18:0 molecule together within the fibrillar crystal, while they are not to break covalent bonds in nanocelluloses. We make the hypothesis that the fibers are broken and partitioned within micron-sized walls, probably being too rigid to bend and guarantee lateral cohesion and connectivity ⊥$Z$. Unfortunately, complete tensile characterization data for individual SLC18:0 fibers and their bundles are not available in the literature and until then these hypotheses are simple speculation.

Lipid lamellar hydrogels are materials seldom described in the literature.[76–78] In their monophasic form, they are generally composed of a homogeneous lipid lamellar phase doped with polymer-grafted lipids or surfactants, of which the role consists in reducing the local bending free energy of the bilayers, thus introducing spurious dislocation defects. Defects are responsible for the propagation of the elastic properties in the macroscopic gel state.[76,77] Alternative biphasic lamellar hydrogels have also been described and constituted by a lipid lamellar phase gelled via an independent low molecular weight gelator.[1,79,80] According to these descriptions, the difference between isotropic fibrillar and anisotropic lamellar hydrogels is very large. In the particular case of GC18:0 lamellar hydrogels, their structure can be described as an interconnected network of defectuous lamellar (of $d$-spacings below 25 nm) domains (of width between 100 μm and 500 μm) surrounded by water. Both the local defects and the heterogeneity in the spatial distribution of the domains seems to be responsible for the mechanical properties of the gel at room temperature (Figure 1).[25,76] A not entirely unrealistic, and thus possible, structure of the lamellar hydrogel before freeze-casting could be described by a spongy solid constituted of lamellar domains of different sizes. After freeze-casting, the ice front ∥Z withdraws water from the lamellar domains, causing a sudden decrease in the *d*-spacing from ~20 nm to ~4 nm. Considering that the thickness of a single GC18:0 bilayer is approximately 3.6 nm,[23,26] the actual water thickness between the bilayers is about 5Å. As a consequence, lamellar domains densify in the foam walls, trapped among the ice crystals, and their orientation change from ⊥Z to ∥Z, through an order-disorder-order transition, visible through *in-situ* 2D SAXS experiments (Figure 4, Figure S 5). As shown by SEM-FEG and CLSM (Figure 2, Videos 1,2 as Supporting Information), the macroscale porosity caused by ice does not follow the expected linear direction of uniaxial freezing, but it is rather confined within the tortuosity of the pre-exiting lamellar phase. Such an isotropic structure is quite atypical for freeze-cast solids. The final isotropic response to compression of the lamellar solid foam is certainly due to the tightly interconnected isotropic lamellar network. Finally, we formulate the hypothesis that the liquid crystalline nature of the GC18:0 lipid layers, compared to the crystalline SLC18:0 fibers, combined with the heterogeneous distribution of the lamellar domains are responsible for their more compliant response upon directional freezing.

These results are summarized in the cartoon illustrated in Figure 6, which shows how the different initial (isotropic) fibrillar or (anisotropic heterogeneous) lamellar gels lead to two macroporous solid foams of different structures and inverted anisotropy. Ice growth probably breaks apart the fibers in the scale range between 10 nm and 100 nm, squeezing them into flat, possibly poorly connected, walls. On the contrary, ice growth only causes disorder in the spatial orientation of the lamellar domains, which, probably due to their liquid crystalline order and hydration, remain fluid and highly interconnected at low temperatures, forming a more isotropic network of self-assembled compressed lipid layers. As a consequence, fibrillar solids are brittle, standing less than 100 times their weight while lamellar spongy foams are able to withstand up to 1000 times their own weight (Figure 5c), making them the first unique case of soft self-assembled foams to have comparable properties with polymer foams. Lastly, we highlight the fact that the entire process is reversible, as one can recover both the fibrillar and lamellar hydrogels by simply rehydration (and sonication) of the foams.

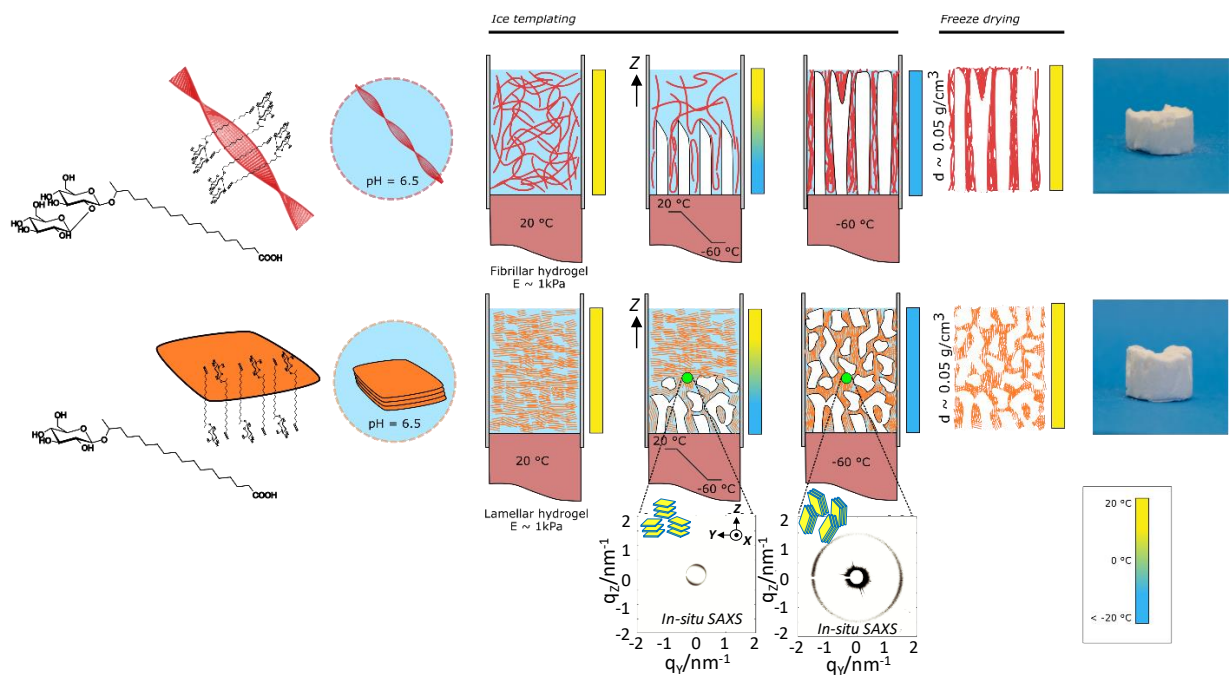

**Figure 6** – Cartoon summarizing the self-assembly, hydrogel-formation process of microbial glycolipids SLC18:0 and GC18:0. Both fibrillar and lamellar hydrogels have comparable elastic moduli before freezing. Freezing from 20°C to -60°C at a typical rate of 5°C/min induces oriented ice crystallization, consequently tilting the orientation of the lamellar phase (SAXS). After ice sublimation, hydrogels have turned into fibrillar (SLC18:0) and lamellar (GC18:0) macroporous monoliths. The fibrillar foam is brittle while the lamellar solid foam can stand up to 1000 times its own weight (Figure 5c).

## Conclusion

This work settles the grounds to cast self-assembled hydrogels into macroporous soft materials at temperatures much below 0°C and under nonequilibrium conditions. We show how self-assembly at the nanoscale and supramolecular arrangement drive the mechanical properties of the solid foams. We employ two chemically-analogous glycolipids, one forming an isotropic self-assembled fibrillar network (SAFiN) hydrogel and the other one a lamellar hydrogel composed of interconnected defectuous lamellar domains, both in the vicinity of pH 6 and with comparable mechanical properties, with $G'$ ~ 0.3-1 kPa. When these hydrogels are processed with unidirectional freeze-casting, with typical freezing rates of 1°C.min$^{-1}$ and 10°C.min$^{-1}$, we find the following.

1) The fibrillar hydrogels systematically form macroporous solids; the pores are oriented in the ice-growth direction (∥Z), as expected for most freeze-cast inorganic and organic solids; the walls are poorly interconnected ⊥$Z$; the walls are composed of tightly packed network of entangled twisted ribbons, that is the typical twisted fibrillar structure is kept intact after freeze-casting. However, we formulate the hypothesis that ice crystallization breaks the fibers at scales above 10 nm confining them in anisotropic, micron-sized, walls. For this reason, these materials exfoliate, they are brittle, highly fragile upon application of a compressive stress ∥$Z$ and their mechanical properties are also highly anisotropic. <span style="color:red">At the moment, we exclude an effect of ice crystallization on disentanglement by analogy to literature data on freeze-cast nanocellulosic foams.</span>

2) The lamellar hydrogels form solid foams; the macroporous structure is spongy and highly interconnected, with low degree of, if any, alignment ∥$Z$, in contrast to most freeze-cast solids; the walls are composed of a dense lamellar phase, of which the orientation tilts from ⊥$Z$ to ∥$Z$ during the freezing process. Ice crystallization induces local disorder in the lamellar domains, but their liquid crystalline nature and hydration are probably responsible for their malleability and structural resistance: the lamellar phase is kept after freeze-casting, but it is more dense (the interlayer water thickness drops from about 16 nm to less than 0.5 nm). This is due to dehydration of the lamellar phase in favour of the growing ice regions during templating. These spongy materials are soft, although quite tough for a single lipid phase, and display an isotropic behaviour to compression, with Young moduli of several kPa, the order of magnitude of living tissues.

In summary, we show that two lipid self-assembled hydrogels similar in composition and of comparable elastic properties, can be cast into macroporous solids with extremely different mechanical properties, and that in tight relationship with the self-assembled nature of the lipid: hydrogels with a isotropic arrangement of fibers yield brittle solids withstanding less than 100 times their weight, while a 3D arrangement of lamellar domains provides soft solid foams resisting up to 1000 times their own weight.

**Acknowledgements**

This work received financial support by the ESRF – The European Synchrotron, Grenoble, France, under the experiment number SC-4479 at the SAXS beamline ID02. We gratefully thank David Montero at Institute des Matériaux de Paris Centre (IMPC) for performing the SEM-FEG experiments. SEF-FEM instrumentation belongs to the facilities of the Institut des Matériaux de Paris Centre (IMPC FR2482) and was funded by UPMC, CNRS and by the C'Nano projects of the Région Ile-de-France. We acknowledge Dr. Sophie Roelants and Prof. Wim Soetaert (Ghent University, Belgium) for providing the monounsaturated sophorolipids and glucolipids and Prof. Chris Stevens (Ghent University, Belgium) for providing access to the hydrogenation device. Maxime Luquiau and Aurélie Grenet are kindly acknowledged for their experimental help. We acknowledge Prof. E. Kontturi (Aalto University, Finland) for providing us a sample of bacterial cellulose, obtained from purified commercial *nata de coco*.
**References:**

1  C. Stubenrauch and F. Gießelmann, *Angew. Chemie - Int. Ed.*, 2016, **55**, 3268–3275.

2  K. L. Scotti and D. C. Dunand, *Prog. Mater. Sci.*, 2014, **94**, 243–305.

3  E. Thibert and F. Dominé, *J. Phys. Chem. B*, 1998, **102**, 4432–4439.

4  Y. Tajima, T. Matsuo and H. Suga, *J. Phys. Chem. Solids*, 1984, **45**, 1135–1144.

5  S. Deville, *Adv. Eng. Mater.*, 2008, **10**, 155–169.

6  U. G. K. Wegst, M. Schecter, A. E. Donius and P. M. Hunger, *Philos. Trans. A. Math. Phys. Eng. Sci.*, 2010, **368**, 2099–121.

7  B. Wicklein, A. Kocjan, G. Salazar-Alvarez, F. Carosio, G. Camino, M. Antonietti and L. Bergström, *Nat. Nanotechnol.*, 2015, **10**, 277–83.

8  D. Chen, Y. Zhang, C. Ni, C. Ma, J. Yin, H. Bai, Y. Luo, F. Huang, T. Xie and Q. Zhao, *Mater. Horizons*, , DOI:10.1039/C9MH00090A.

9  J. Yu, M. Borghei, Y. Fan, Z. Wang, L. Bai, L. Liu, O. J. Rojas and A. Tripathi, *ACS Nano*, , DOI:10.1021/acsnano.8b07235.

10  S. Deville, *J. Mater. Res.*, 2013, **28**, 2202–2219.

11  P. A. Albouy, S. Deville, A. Fulkar, K. Hakouk, M. Impéror-Clerc, M. Klotz, Q. Liu, M. Marcellini and J. Perez, *Soft Matter*, 2017, **13**, 1759–1763.

12  M. L. Ferrer, R. Esquembre, I. Ortega, C. R. Mateo and F. Monte, *Chem. Mater.*, 2006, **18**, 554–559.

13  J. Dhainaut, G. Piana, S. Deville, C. Guizard and M. Klotz, *Chem. Commun.*, 2014, **50**, 12572–12574.

14  X. Du, J. Zhou, J. Shi and B. Xu, *Chem. Rev.*, 2015, **115**, 13165–13307.

15  N. F. Goldshleger, A. S. Lobach, V. E. Baulin and A. Y. Tsivadze, *Russ. Chem. Rev.*,

2017, **86**, 269–297.

16  K. Y. Lee and D. J. Mooney, *Chem. Rev.*, 2001, **101**, 203–225.

17  R. Yoshida and T. Okano, in *Biomedical Applications of Hydrogels Handbook*, eds. R. M. Ottenbrite, K. Park and T. Okano, Spinger, 2010, vol. c, pp. 19–43.

18  J. Raeburn, A. Zamith Cardoso and D. J. Adams, *Chem. Soc. Rev.*, 2013, **42**, 5143–5156.

19  M. C. Nolan, A. M. Fuentes Caparrós, B. Dietrich, M. Barrow, E. R. Cross, M. Bleuel, S. M. King and D. J. Adams, *Soft Matter*, 2017, **13**, 8426–8432.

20  D. Berillo, B. Mattiasson, I. Yu and H. Kirsebom, *J. Colloid Interface Sci.*, 2012, **368**, 226–230.

21  A. S. Cuvier, F. Babonneau, J. Berton, C. V. Stevens, G. C. Fadda, I. Genois, P. Le Griel, G. Péhau-Arnaudet and N. Baccile, *Chem. - An Asian J.*, 2015, **10**, 2419–2426.

22  A.-S. Cuvier, J. Berton, C. V Stevens, G. C. Fadda, F. Babonneau, I. N. a Van Bogaert, W. Soetaert, G. Pehau-Arnaudet and N. Baccile, *Soft Matter*, 2014, **10**, 3950–9.

23  N. Baccile, M. Selmane, P. Le Griel, S. Prévost, J. Perez, C. V. Stevens, E. Delbeke, S. Zibek, M. Guenther, W. Soetaert, I. N. A. Van Bogaert and S. Roelants, *Langmuir*, 2016, **32**, 6343–6359.

24  G. Ben Messaoud, P. Le Griel, D. Hermida-Merino, S. L. K. W. Roelants, W. Soetaert, C. V. Stevens and N. Baccile, *Chem. Mater.*, 2019, DOI: 10.1021/acs.chemmater.9b01230.

25  G. Ben Messaoud, P. Le Griel, S. Prévost, D. H. Merino, W. Soetaert, S. L. K. W. Roelants, C. V. Stevens and N. Baccile, *arXiv*, 2019, 1907.02223.

26  N. Baccile, A.-S. Cuvier, S. Prévost, C. V Stevens, E. Delbeke, J. Berton, W. Soetaert, I. N. A. Van Bogaert and S. Roelants, *Langmuir*, 2016, **32**, 10881–10894.

27  P.-G. de Gennes, *Macromolecules*, 1976, **9**, 587–593.

28  R. de Rooij, D. van den Ende, M. H. G. Duits and J. Mellema, *Phys. Rev. E*, 1994, **49**, 3038–3049.

29  P. J. Flory, *Principles of polymer chemistry*, Cornell University Press, Ithaca, NY, NY, 1953.

30  S. Q. Chen, P. Lopez-Sanchez, D. Wang, D. Mikkelsen and M. J. Gidley, *Food Hydrocoll.*, 2018, **81**, 87–95.

31  M. A. Greenfield, J. R. Hoffman, M. O. De La Cruz and S. I. Stupp, *Langmuir*, 2010, **26**, 3641–3647.

32  N. Baccile, L. Van Renterghem, P. Le Griel, G. Ducouret, M. Brennich, V. Cristiglio, S. L. K. W. Roelants and W. Soetaert, *Soft Matter*, 2018, **14**, 7859–7872.

33  C. Colquhoun, E. R. Draper, R. Schweins, M. Marcello, D. Vadukul, L. C. Serpell and D. J. Adams, *Soft Matter*, 2017, **13**, 1914–1919.

34  A. Jaishankar and G. H. McKinley, *Proc. R. Soc. A Math. Phys. Eng. Sci.*, , DOI:10.1098/rspa.2012.0284.

35  F. Gobeaux, E. Belamie, G. Mosser, P. Davidson and S. Asnacios, *Soft Matter*, 2010, **6**, 3769–3777.

36  H. H. Winter and F. Chambon, *J. Rheol. (N. Y. N. Y).*, 1986, **30**, 367.

37  P. Sollich, F. Lequeux, P. Hébraud and M. E. Cates, *Phys. Rev. Lett.*, 1997, **78**, 2020–2023.

38  P. Sollich, *Phys. Rev. E - Stat. Physics, Plasmas, Fluids, Relat. Interdiscip. Top.*, 1998, **58**, 738–759.

39  H. J. Hwang, R. A. Riggleman and J. C. Crocker, *Nat. Mater.*, 2016, **15**, 1031–1036.

40  S. Christoph, A. Hamraoui, E. Bonnin, C. Garnier, T. Coradin and F. M. Fernandes, *Chem. Eng. J.*, 2018, **350**, 20–28.

41  N. L. Francis, P. M. Hunger, A. E. Donius, B. W. Riblett, A. Zavaliangos, U. G. K. Wegst and M. a. Wheatley, *J. Biomed. Mater. Res. - Part A*, 2013, **101**, 3493–3503.

42  N. Lavoine and L. Bergström, *J. Mater. Chem. A*, 2017, **5**, 16105–16117.

43  P. Munier, K. Gordeyeva, L. Bergström and A. B. Fall, *Biomacromolecules*, 2016, **17**, 1875–1881.

44  T. Dvir, L. Fink, R. Asor, Y. Schilt, A. Steinar and U. Raviv, *Soft Matter*, 2013, **9**, 10640.

45  L. Fink, J. Feitelson, R. Noff, T. Dvir, C. Tamburu and U. Raviv, *Langmuir*, 2017, **33**, 5636–5641.

46  G. Brotons, M. Dubois, L. Belloni, I. Grillo, T. Narayanan and T. Zemb, *J. Chem. Phys.*, 2005, **123**, 024704.

47  O. Lotan, L. Fink, A. Shemesh, C. Tamburu and U. Raviv, *J. Phys. Chem. A*, 2016, **120**, 3390–3396.

48  P. Versluis, J. C. Van de Pas and J. Mellema, *Langmuir*, 2001, **17**, 4825–4835.

49  N. Baccile, R. Noiville, L. Stievano and I. Van Bogaert, *Phys. Chem. Chem. Phys.*, 2013, **15**, 1606–1620.

50  J. Blackwell, P. D. Vasko and J. L. Koenig, *J. Appl. Phys.*, 1970, **41**, 4375–4379.

51  J. Oomens and J. D. Steill, *J. Phys. Chem. A*, 2008, **112**, 3281–3283.

52  R. H. Marchessault, *Pure Appl. Chem.*, 1962, **5**, 107–129.

53  G. Porte, J. Marignan, P. Bassereau and R. May, *Eur. Lett.*, 1988, **7**, 713–717.

54  D. Roux and C. Safinya, *J. Phys.*, 1988, **49**, 307–318.

55  B. Demé, M. Dubois, T. Gulik-krzywicki and T. Zemb, *Lagmuir*, 2002, **18**, 997–1004.

56  B. Demé, M. Dubois and T. Zemb, *Langmuir*, 2002, **18**, 1005–1013.

57  M. Dubois, T. Zemb, L. Belloni, A. Delville, P. Levitz and R. Setton, *J. Chem. Phys.*, 1992, **96**, 2278–2286.

58  R. M. Pashley and J. N. Israelachvili, *J. Colloid Interface Sci.*, 1984, **101**, 511–523.

59  T. J. McIntosh and S. A. Simon, *Annu. Rev. Biophys. Biomol. Struct.*, 1994, **23**, 27–51.

60  R. P. Rand and V. A. Parsegian, *BBA - Rev. Biomembr.*, 1989, **988**, 351–376.

61  M. Dubois, T. Zemb, N. Fuller, R. P. Rand and V. A. Parsegian, *J. Chem. Phys.*, 1998, **108**, 7855–7869.

62  B. Thomas, N. Baccile, S. Masse, C. Rondel, I. Alric, R. Valentin, Z. Mouloungui, F. Babonneau and T. Coradin, *J. Sol-Gel Sci. Technol.*, 2011, **58**, 170–174.

63  M. Kanduč, A. Schlaich, A. H. de Vries, J. Jouhet, E. Maréchal, B. Demé, R. R. Netz and E. Schneck, *Nat. Commun.*, 2017, **8**, 14899.

64  F. Ricoul, M. Dubois, L. Belloni, T. Zemb, C. André-Barrès and I. Rico-Lattes, *Langmuir*, 1998, **14**, 2645–2655.

65  C. Yu, J. Bianco, C. Brown, L. Fuetterer, J. F. Watkins, A. Samani and L. E. Flynn, *Biomaterials*, 2013, **34**, 3290–3302.

66  A. Samani, J. Zubovits and D. Plewes, *Phys. Med. Biol.*, 2007, **52**, 1565--1576.

67  M. F. Ashby, L. J. Gibson, U. Wegst and R. Olive, *Proc. R. Soc. A Math. Phys. Eng. Sci.*, 1995, **450**, 123–140.

68  S. Budday, R. Nay, R. De Rooij, P. Steinmann, T. Wyrobek, T. C. Ovaert and E. Kuhl, *J. Mech. Behav. Biomed. Mater.*, 2015, **46**, 318–330.

69  B. A. M. Kloxin, C. J. Kloxin, C. N. Bowman and K. S. Anseth, *Adv. Mater.*, 2010, **22**, 3484–3494.

70  R. Yu, N. Lin, W. Yu and X. Y. Liu, *CrystEngComm*, 2015, **17**, 7986–8010.

71  H. Sehaqui, M. Salajková, Q. Zhou and L. A. Berglund, *Soft Matter*, 2010, **6**, 1824–1832.

72  J. Lee and Y. Deng, *Soft Matter*, 2011, **7**, 6034–6040.

73  P. Srinivasa and A. Kulachenko, *Mech. Mater.*, 2015, **80**, 13–26.

74  X. Wang, Y. Zhang, C. Zhi, X. Wang, D. Tang, Y. Xu, Q. Weng, X. Jiang, M. Mitome, D. Golberg and Y. Bando, *Nat. Commun.*, 2013, **4**, 2905.

75  K. Suresh, S. Patil, P. Ramanpillai Rajamohanan and G. Kumaraswamy, *Langmuir*, 2016, **32**, 11623–11630.

76  H. E. Warriner, S. H. Idziak, N. L. Slack, P. Davidson and C. R. Safinya, *Science (80-. ).*, 1996, **271**, 969–73.

77  J. Niu, D. Wang, H. Qin, X. Xiong, P. Tan, Y. Li, R. Liu, X. Lu, J. Wu, T. Zhang, W. Ni and J. Jin, *Nat. Commun.*, 2014, **5**, 3313.

78  N. L. Slack, M. Schellhorn, P. Eiselt, M. a. Chibbaro, U. Schulze, H. E. Warriner, P. Davidson, H.-W. Schmidt and C. R. Safinya, *Macromolecules*, 1998, **31**, 8503–8508.

79  K. Steck, J. H. van Esch, D. K. Smith and C. Stubenrauch, *Soft Matter*, , DOI:10.1039/c8sm02330a.

80  S. Koitani, S. Dieterich, N. Preisig, K. Aramaki and C. Stubenrauch, *Langmuir*, , DOI:10.1021/acs.langmuir.7b02101.

Supporting Information

# Soft lamellar solid foams from ice-templating of self-assembled lipid hydrogels: organization drives the mechanical properties

**Niki Baccile,**[a*] **Ghazi Ben Messaoud,**[a,†] **Thomas Zinn,**[b] **Francisco Fernandes**[a,*]

[a] Sorbonne Université, Centre National de la Recherche Scientifique, Laboratoire de Chimie de la Matière Condensée de Paris, LCMCP, F-75005 Paris, France

[†] Current address: DWI- Leibniz Institute for Interactive Materials, Forckenbeckstrasse 50, 52056 Aachen, Germany

[b] ESRF - The European Synchrotron, 71 Avenue des Martyrs, 38043 Grenoble, France

**Materials and Methods**

*Products.* Acidic deacetylated C18:0 sophorolipids (SLC18:0) and acidic deacetylated C18:0 glucolipids (GC18:0) have been used from previously existing batch samples, the preparation and characterization of which is published elsewhere.[1,2] Acid (HCl 37%) and base (NaOH) are purchased at Aldrich. MilliQ-quality water has been employed throughout the experimental process. 18:1 Liss Rhod PE (Liss) ($M_w$= 1301.7 g.mol$^{-1}$, $\lambda_{abs}$= 560 nm, $\lambda_{em}$= 583 nm), 1,2-dioleoyl-sn-glycero-3-phosphoethanolamine-N-(lissamine rhodamine B sulfonyl) (ammonium salt), is purchased by Avanti® Polar, Inc.

*Self-assembly and preparation of hydrogels.* GC18:0 and SLC18:0 are glycolipids respectively constituted by a β-D-glucose and sophorose (D-glucose β(1,2)) headgroup linked to stearic acid via a glyosidic bond located on the C17 carbon of stearic acid (Figure 1). Their self-assembly properties to form bilayers (GC18:0) and twisted flat ribbons (SLC18:0) were reported by us.[1,2,3] Briefly, the selected molecule is introduced in water at the indicated concentration (please refer to the main text for the exact values), and to promote solubility, the pH is raised at values between 10 and 11, at which both compounds mainly form micelles. pH is then reduced to about pH 6.2 ± 0.3 according to the procedure published elsewhere.[4,5] Fibrillar SLC18:0 gels are nicely obtained by a controlled (< 50 μL/h) acidification using 0.5 M or 1 M HCl solutions.[5] Slow acidification is crucial to obtain a gel, as explained in ref. 5. Acidification of lamellar hydrogels can also be carried out by hand but intercalating sonication and vigorous vortexing during each acid addition below pH 8, otherwise local lamellar aggregates form. An alternative, highly reproducible, procedure to prepare GC18:0 lamellar gels was detailed in ref. 4 and consists in simply dispersing the smaple in water, followed by sonication and adjustment of pH to 6.2 and ionic strength to about 100 mM. Sonication (10-15 min) and gentle heating at 70°C during less than 5 min followed by cooling provides a stable gel within less than 2 h. For the GC18:0 system, the method of reaching a final pH of 6.2 is not important, as long as solution is homogeneous and the total ionic strength below 100 mM. Both hydrogels can be processed as such after their preparation.

*Solid foams by freeze-casting.* The unidirectional freeze-casting setup is home-built according to the literature.[6,7] The setup consists in a liquid nitrogen Dewar, a 40 cm copper bar (Ø¼1.5 cm), a heating element and a polypropylene tube partially inserted in the hot end of the copper bar to hold the sample prior to freezing. A silicon mold (1 cm x 1 cm x 1 cm) sitting on top of the copper bar was occasionally used for those experiments requiring longitudinal and transverse analysis of the mechanical properties, as clearly indicated in the legends of the corresponding figures in the main text. *Note: The direction of freeze-casting, that is of the ice-growing front, is identified as the Z-axis throughout the paper. Nomenclatures ∥Z and ⊥Z respectively refer to longitudinal and transversal directions with respect to the ice-growing front.*

The assembly was carried out in such a manner that half of the copper bar plunges into liquid $N_2$ to create a heat sink. The temperature of the opposed extremity of the copper was controlled by the simultaneous action of the heat sink and the heating element. The heating element was controlled by a dedicated PID thermocontroller able to modulate the cooling rate between 1°C.min$^{-1}$ and 10°C.min$^{-1}$. A temperature sensor (K thermocouple) is located at the bottom of the cell, close to the tip of the copper bar. In a typical experiment, 2 mL of the selected self-assembled glycolipid sample (in solution or hydrogel state) is poured inside the polypropylene tube, in direct contact with the copper surface. After a 5 min equilibration time at 20°C, the sample is cooled down to -60°C, removed from the setup and placed at -20°C before freeze drying. Ice sublimation is systematically conducted on all freeze-cast samples in a Christ Alpha 2-4 LD freeze dryer. The temperature of the freeze dryer condenser is kept below -60°C and the internal pressure stabilized within few minutes to approximately 5 x 10$^{-5}$ bar. The freeze drying process is left to proceed for 24 h, allowing for the recovery of a dried lightweight, macroporous, soft solid, which will be referred to as a solid foam throughout this work.

*Small Angle X-ray Scattering (SAXS)*: SAXS experiments have been performed on the ID02 beamline at the ESRF synchrotron facility (Grenoble, France). The experiments have been done at 17.0 keV and two sample-to-detector distances were used: 2 m and 8 m. Calibration of the q-range is done using silver behenate as classical standard ($d_{ref}$ = 58.38 Å). The signal of the CCD camera, used to record the data, is normalized and integrated azimuthally.[8] to obtain the typical one-dimensional scattering profile I(q) i.e. scattered intensity versu scattering vector q, where q is given by $4\pi/\lambda \sin(\theta)$ with the scattering angle $2\theta$. The scattered intensity I(q) is given in the dimensionless units of sr$^{-1}$. I(q) data are presented as such and have not been corrected for the empty and water cell has been performed. Typical acquisition times were in the order of 100 ms, which we considered enough to obtain a good signal-to-noise profile, with no beam damage observed. One spectrum per temperature value or per time is recorded.

The typical freeze-casting experiment has been adapted to the beamline via a home-made freeze-casting device, which exposes a 2 mm flat cell to beam, whereas the cell is supported by a plastic holder containing two face-to-face Kapton© windows. The image and scheme of the device are shown in Figure S 3a,b. The heating element is controlled by a dedicated PID thermocontroller able to modulate the cooling rate between 1°C.min$^{-1}$ and 10°C.min$^{-1}$. Temperature is hence controlled between +20°C and -60°C. The SAXS patterns are recorded at a step of 5°C in the entire temperature range at different positions in the cell: for each temperature, the signal is collected at five positions (Z-axis) simultaneously, namely, 100 µm (the closest to the bar), 500 µm, 900 µm, 1300 µm, 1700 µm from the top of the copper bar (Figure S 3b). Movement of the stage along Z is controlled by an automated stage available at the beamline. Both the acquisition time and the stage displacement are fast enough (~ second, including signal acquisition and displacement for the five positions) with respect to the cooling rate, which is 0.17 °C.s$^{-1}$ for the fastest rate. In view of these considerations, one can consider that the measurement can be considered instantaneous (the sample is in the same physical state for all positions at a given measurement time) for all positions.

*Scanning Electron Microscopy with Field Emission Gun (SEM-FEG)*: SEM-FEG experiments have been recorded on a Hitachi SU-70. The images were taken in secondary electron mode with an accelerating voltage at 1 kV, 5 kV or 10 kV. Prior to analysis, the materials were coated with a thin layer of gold by sputter deposition. Both the ∥Z and ⊥Z sections of the samples were systematically observed. SEM images have been treated using the software Image-J,[9] and in particular the local roughness analysis has been done using the roughness calculation plugin, freely available and implemented for ImageJ. The local roughness analysis is measured on the whole surface and it gives roughness values according to the ISO 4287/2000 standard. We present the root mean square roughness, $R_{RMS} = \sqrt{\frac{1}{n}\sum_{i=1}^{n} y_i^2}$, where the roughness profile contains $n$ equally spaced points along the trace, and where $y_i$ is the vertical distance from the mean line to the i$^{th}$ data point. Height is assumed to be positive in the up direction, away from the bulk material.

*X-Ray Diffraction (XRD)*: XRD experiments have been recorded on a Bruker D8 Advance using a classical Bragg Brentano θ-2θ configuration. A copper K$_\alpha$ anticathode with a wavelength λ= 0.154 nm is used. A 1D LynxEye detector is used. We have employed a 0.05° step and 0.5 s/step as acquisition parameters.

*Fourier Transform InfraRed spectroscopy (FTIR)*: FTIR experiments have been done in the ATR mode using a Perkin Elmer Spectrum 400 instrument.

*Rheology:* Viscoelastic measurements were carried out using an Anton Paar MCR 302 rheometer equipped with parallel titanium or stainless steel sandblasted plates (diameter 25 mm). All experiments were conducted at 25 °C and the temperature was controlled by the stainless steel lower plate, which is the surface of the Peltier system. During experiments, the measuring geometry was covered with a humidity chamber to minimize water evaporation. To characterize the hydrogels, strain sweep experiments were first conducted by changing the shear strain (γ) from 0.001% to 100% to determine the linear viscoelastic regime (LVR). After loading a new sample, values between γ = 0.02 – 0.05 % within the LVR were used in the subsequent angular frequency sweep from ω = 100 and 0.01 rad.s$^{-1}$.

*Mechanical analysis*: the mechanical properties of the freeze-cast matrices were carried out on an Anton Paar MCR 302 rheometer equipped with a force transducer of 50 N and a plate-plate (25 mm) geometry. The freeze-cast foams obtained from SLC18:0 being friable, the uniaxial compression experiments were conducted only on GC18:0 samples prepared at a freezing rate of 5°C.min$^{-1}$. The foams were compressed along the Z-axis from 10 mm to 0.2 mm with a linear compression speed of 5 µm.s$^{-1}$. The gap ($l$) and the normal force ($F$) being imposed were measured simultaneously at the upper plate. The force-displacement responses were re-plotted in terms of stress ($\sigma$) and strain ($\varepsilon$), with $\varepsilon(\%) = \frac{l_0-l}{l_0} * 100$, where $l_0$ (m) is the original height of the foam and $l$ (m) is the current height during compression; $\sigma(kPa) = \frac{F}{A*1000}$, where $A$ (m$^2$) is the cross-sectional area of the foams.

*Confocal Laser Scanning Microscopy (CLSM)*: CLSM was performed with a LeicaSP8 Tandem Confocal system. Samples were excited with the dye specific wavelength (561 nm) and the emission was detected between 580 and 620 nm using a photomultiplier tube (PMT) detector. CLSM images were analyzed using FIJI[9] and 3D construction and projection from Z-stacks was performed using the 3D visualization module of the Leica Application Suite X (LAS X) software. The GC18:0 lamellar solid foam for the CLSM experiments was prepared as follows: a volume of 4 μL of an ethanolic solution of 18:1 Liss Rhod PE (C= 53 mg/mL) was added to 1.5 mL of the GC18:0 hydrogel ($C_{G-C18:0}$= 2.5 wt%, pH 6) to reach an approximate molar ratio of G-C18:0/Liss of 500. Liss is a water insoluble, rhodamine-containing, lipid and it is largely used to mark lipid bilayers. It is generally considered not to interfere with the bilayer assembly at Lipid/Liss ratio above 200. We did not observe any variation in the gel physical aspect after addition of Liss. The gel was freeze-cast at a rate of 5°C/min in a from 20°C to -60°C, freeze-dried et analyzed. The foams were analyzed both ⊥Z and ∥Z, respectively the orthogonal and longitudinal planes with respect to the ice-growth direction (Z-axis).

Orientation analysis of the foams structures was performed ∥Z on sequential Z planes obtained from CLSM volumetric data (videos 1 and 4). Fourier components were computed and binned between -90° < φ < 90° using the directionality tool on FIJI software.

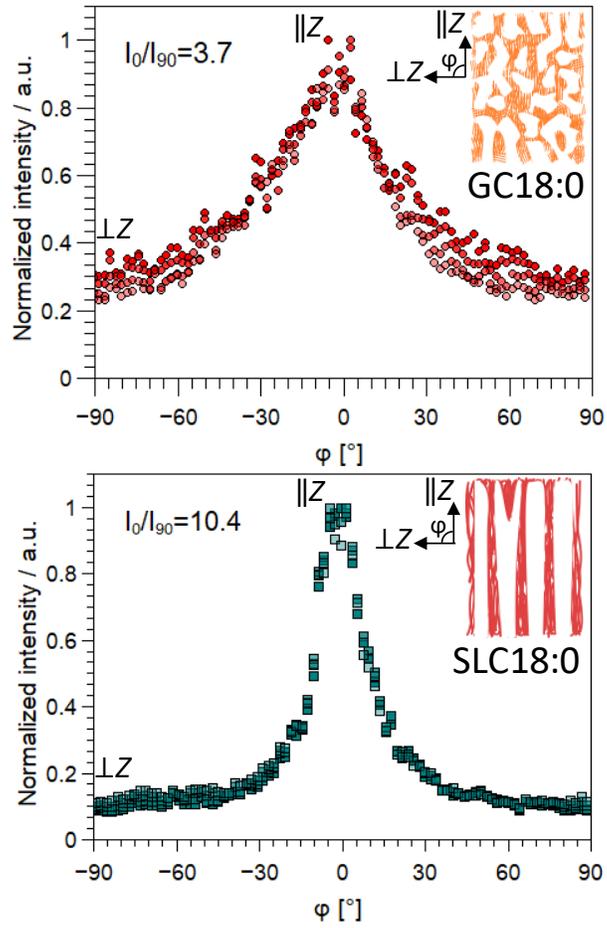

**Figure S 1 – Orientational distribution of the intensity measured on four sequential planes obtained form volumetric CSLM data presented on Video 1 and Video 4, respectively for lamellar GC18:0 and fibrillar SLC18:0 solid foams.**

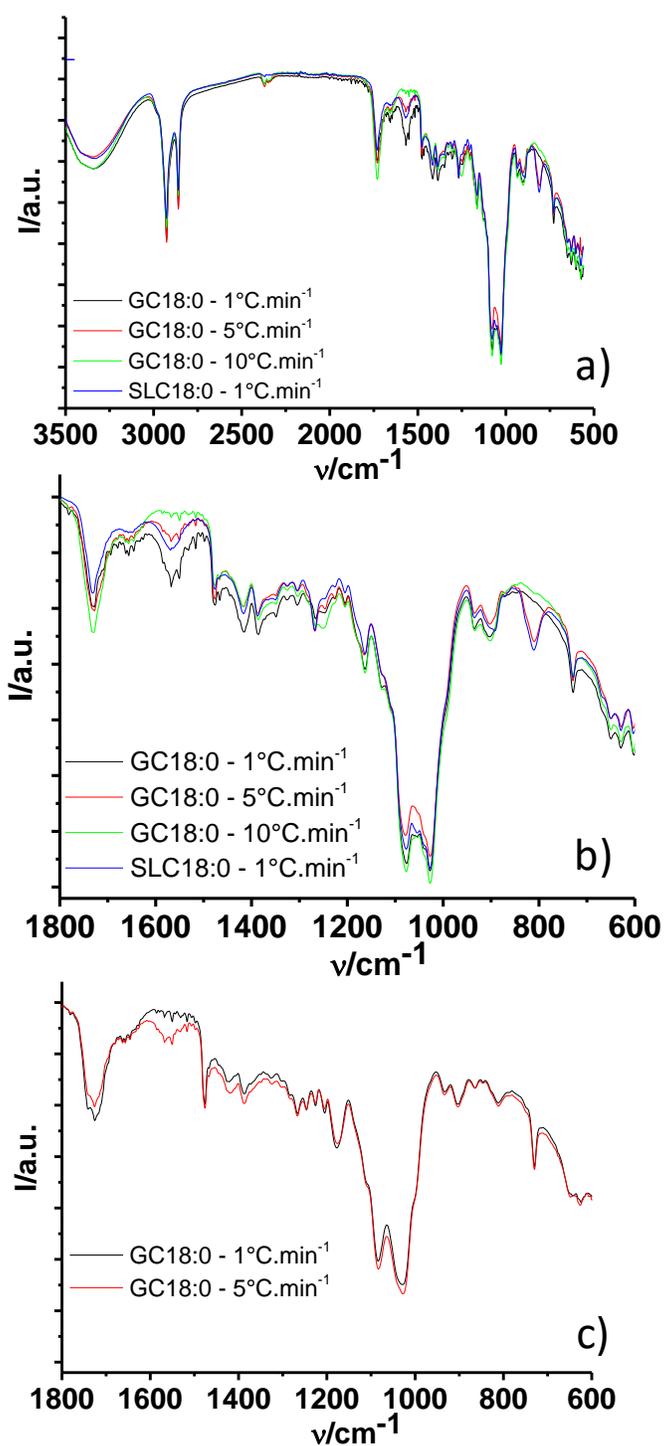

**Figure S 2 – a)** FT-IR experiments performed on SLC18:0 and GC18:0 solid foams freeze-cast at various freezing rates. **b-c)** Highlight of the 1800 – 600 cm$^{-1}$ region of the FT-IR spctrum and zoom on the GC18:0 sample alone (c)

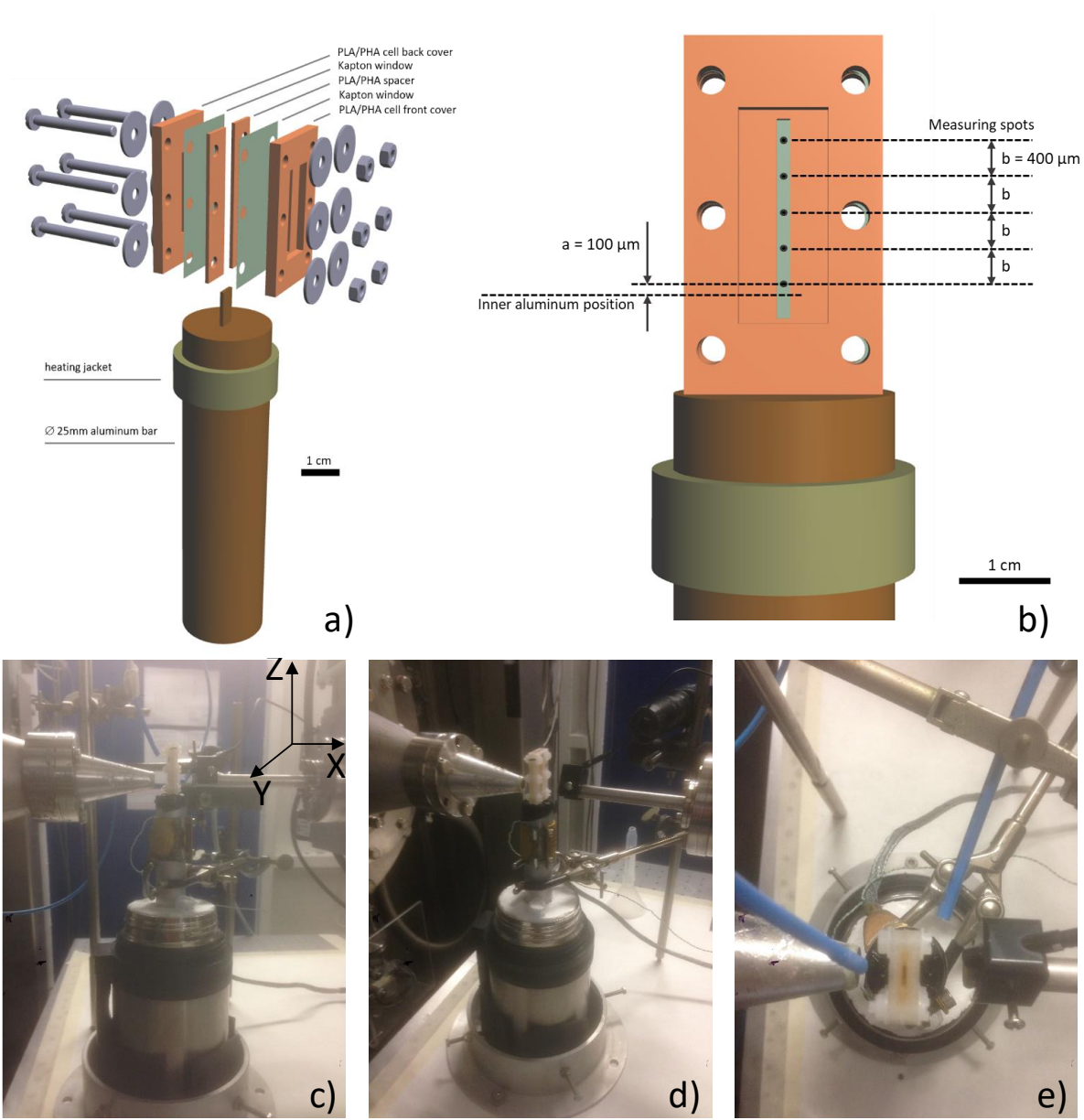

**Figure S 3** – a) Freeze-casting setup used in synchrotron experiments. The bottom part of the aluminum bar (not shown) is kept in liquid nitrogen. The heating jacket is controlled by a dedicated PID and the temperature sensor (K thermocouple) is located at the top of the aluminum bar. b) Detail of the *in situ* freeze-casting cell used in synchrotron experiments. The cell is assembled from 3D printed PVA/PHA parts and kapton tape, assembled by 6 M6 nylon screws and knobs. The measuring position spots within the cell are indicated by black dots. c-e) Side, front and top views of the freeze-casting cell coupled to the ID02 beamline at ESRF synchrotron. The liquid nitrogen Dewar is located at the bottom of the cell is while the X-ray source is on the right-hand side. The blue pipes in (e) carry a constant air flux in the front and bottom of the cell to avoid condensation and crystallization of moisture.

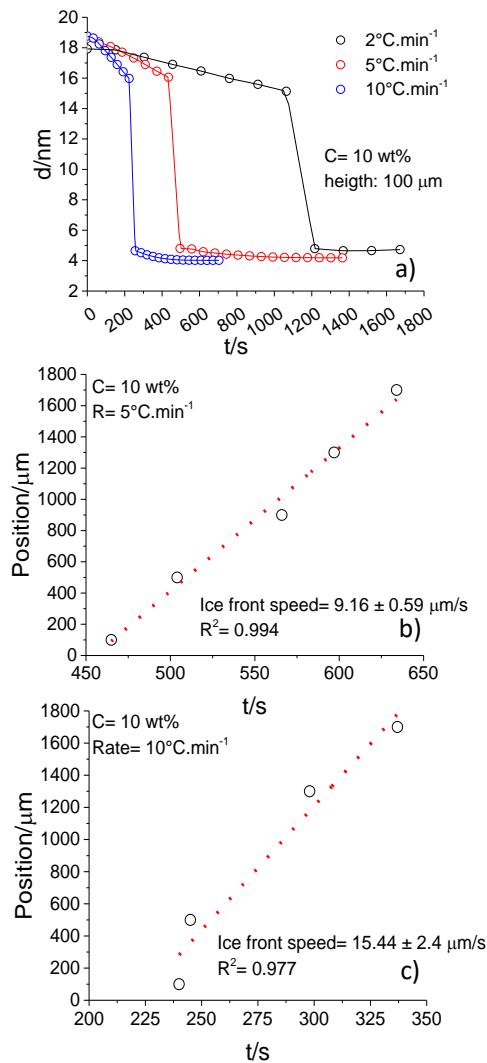

**Figure S 4** – a) Raw data showing the evolution of lamellar *d*-spacing (extracted from *in situ* SAXS) with time measured for GC18:0 hydrogel (C= 10 wt%) at the position of 100 µm from the metal bar (Figure S 3b) for various freezing rates. The sudden drop in the *d*-spacing occurs during freezing and it is assumed to be the ice front. b-c) Evolution of the ice front at positions 100, 500, 900, 1300, 1700 µm from the metal bar (Figure S 3b) with time for the same GC18:0 hydrogel (C= 10 wt%) and two freezing rates (5°C.min$^{-1}$ and 10°C.min$^{-1}$). Data are fitted linearly to estimate the ice front speed. By knowing the measuring position (Figure S 3b), the freezing rate and the ice front speed at a given rate, it is possible to precisely estimate the evolution of *d*-spacing with temperature at a given position and freezing rate. These data are shown in Figure 4c,d in the main manuscript. This calibration is necessary due to the large differences in terms of distance between position at 100 µm (where temperature is actually measured using a thermocouple) and 1700 µm.

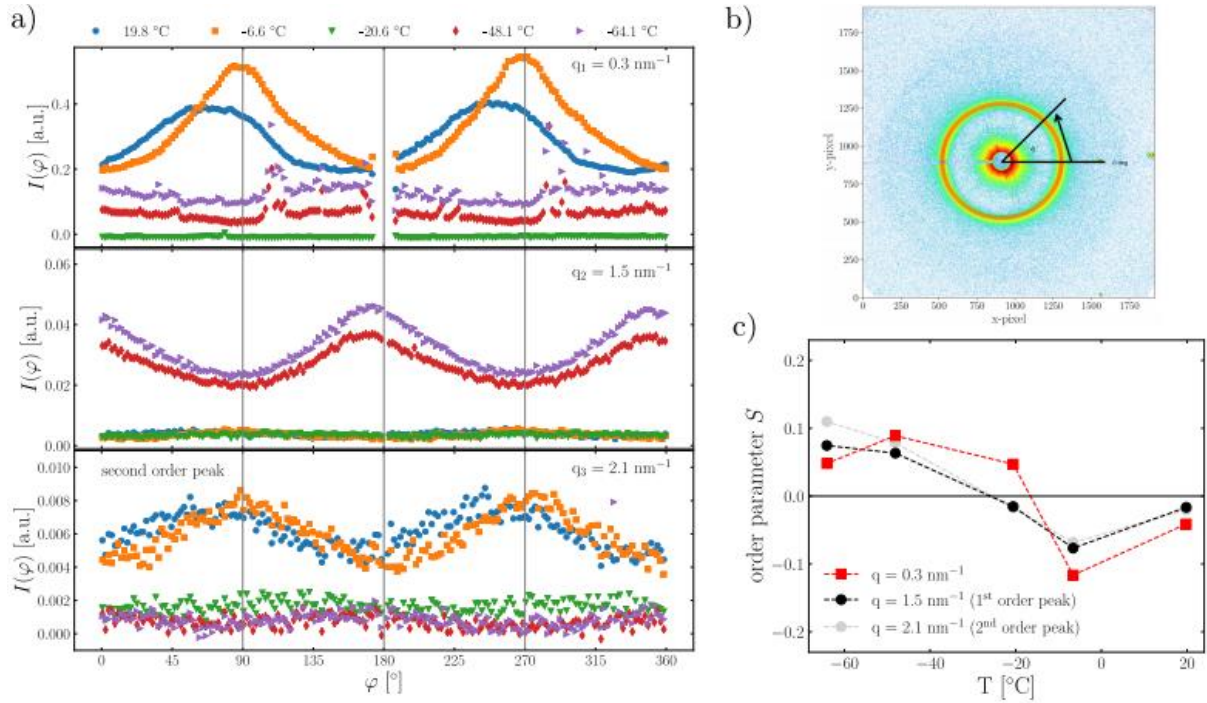

**Figure S 5** - a) Azimuthal intensity distribution for the observed peaks at q = 0.3 nm$^{-1}$, q = 1.5 nm$^{-1}$ and its second order peak at q = 2.1 nm$^{-1}$, b) 2D scattering pattern at T = -64.1°C with the indication of the φ-integration direction and c) obtained order parameter S as a function of temperature for the different peaks.

In Figure S 5a) the azimuthal intensity distribution on the observed peaks is shown as a function of azimuthal angle φ and temperature. The direction of azimuthal integration is depicted in Figure S 5b). For the higher temperatures one can also observe the second order peak at $q_3$=2.1 nm$^{-1}$. Hence, $q_1$ and $q_3$ have similar features i.e. orientation. Whereas $q_2$ appears from -20.6°C to the lowest measured temperature having a 90° shift in orientation. The peaks above -20.6°C completely disappeared. Having the data shown in Figure S 5a) one can easily calculate the orientation order parameter S for the different q values and temperatures according to

$$S_q = \langle P_2(cos\varphi) \rangle_q = \frac{3\langle cos^2\varphi \rangle_q - 1}{2}$$

with $\langle cos^2\varphi \rangle_q = \frac{\int_0^{\pi/2} I(q,\varphi)\, cos^2\varphi\, sin\varphi\, d\varphi}{\int_0^{\pi/2} I(q,\varphi)\, sin\varphi\, d\varphi}$. The obtained S is plotted in Figure S 5c as a function of temperature.

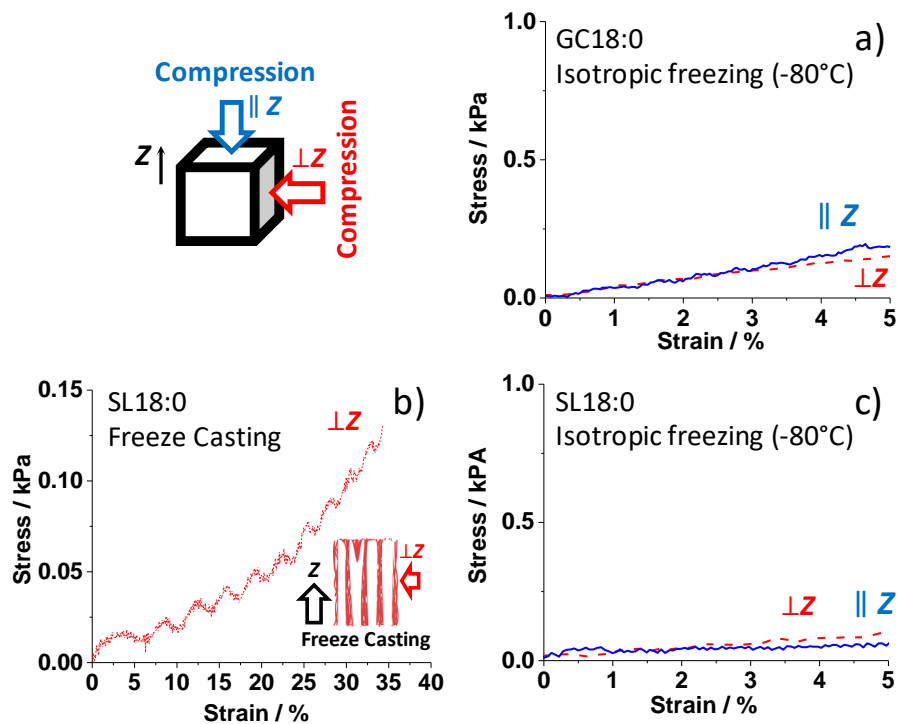

**Figure S 6** – Stress–strain curves of SLC18:0 fibrillar and GC18:0 lamellar foams at 5 wt% measured on cubic samples (1 cm x 1 cm x 1 cm) by compression experiments along ∥Z (blue curves) and ⊥Z (red curves) according to the scheme. Experiments in a) and c) are recorded on samples, that are frozen in an isotropic environment at -80°C. Experiment in b) highlights a broad range of strain values (⊥Z) for the freeze-casting experiment shown in Figure 5b.

[1] A.-S. Cuvier, J. Berton, C. V Stevens, G. C. Fadda, F. Babonneau, I. N. a Van Bogaert, W. Soetaert, G. Pehau-Arnaudet and N. Baccile, *Soft Matter*, 2014, 10, 3950–9

[2] N. Baccile, M. Selmane, P. Le Griel, S. Prévost, J. Perez, C. V. Stevens, E. Delbeke, S. Zibek, M. Guenther, W. Soetaert, I. N. A. Van Bogaert and S. Roelants, *Langmuir*, 2016, 32, 6343–6359

[3] N. Baccile, A.-S. Cuvier, S. Prévost, C. V Stevens, E. Delbeke, J. Berton, W. Soetaert, I. N. A. Van Bogaert and S. Roelants, *Langmuir*, 2016, 32, 10881–10894.

[4] G. Ben Messaoud, P. Le Griel, S. Prévost, D. H. Merino, W. Soetaert, S. L. K. W. Roelants, C. V. Stevens and N. Baccile, *arXiv*, 2019, 1907.02223.

[5] G. Ben Messaoud, P. Le Griel, D. H. Merino, S. L. K. W. Roelants, C. V. Stevens and N. Baccile, *Chem. Mater.,* 2019, DOI: 10.1021/acs.chemmater.9b01230.

[6] S. Christoph, J. Kwiatoszynski, T. Coradin and F. M. Fernandes, *Macromol. Biosci.*, 2016, 16, 182–187.

[7] U. G. K. Wegst, M. Schecter, A. E. Donius and P. M. Hunger, *Philos. Trans. R. Soc. A Math. Phys. Eng. Sci.*, 2010, 368, 2099–2121.

[8] T. Narayanan, M. Sztucki, P. Van Vaerenbergh, J. Leonardon, J. Gorini, L. Claustre, F. Sever, J. Morse, and P. Boesecke. A multipurpose instrument for time-resolved ultra-small-angle and coherent X-ray scattering, *J. Appl. Crystallogr.*, 2018, 51 (Pt 6), 1511-1524

[9] C. A. Schneider, W. S. Rasband and K. W. Eliceiri, *Nat. Methods*, 2012, 9, 671–675.